\documentclass[a4paper,12pt,numbers,sort&compress]{article}
\usepackage[paper=letterpaper]{geometry}

\usepackage{amsmath,amssymb,amsfonts,epsfig,cite,bigstrut,varwidth}
\pdfoutput=1

\begin{document}


\vskip 0.25in

\newcommand{\todo}[1]{{\bf ?????!!!! #1 ?????!!!!}\marginpar{$\Longleftarrow$}}
\newcommand{\fref}[1]{Figure~\ref{#1}}
\newcommand{\tref}[1]{Table~\ref{#1}}
\newcommand{\sref}[1]{\S~\ref{#1}}
\newcommand{\nn}{\nonumber}
\newcommand{\tr}{\mathop{\rm Tr}}
\newcommand{\comment}[1]{}

\newcommand{\cM}{{\cal M}}
\newcommand{\cI}{{\cal I}}
\newcommand{\cW}{{\cal W}}
\newcommand{\cN}{{\cal N}}
\newcommand{\cH}{{\cal H}}
\newcommand{\cK}{{\cal K}}
\newcommand{\cZ}{{\cal Z}}
\newcommand{\cO}{{\cal O}}
\newcommand{\cB}{{\cal B}}
\newcommand{\cC}{{\cal C}}
\newcommand{\cD}{{\cal D}}
\newcommand{\cE}{{\cal E}}
\newcommand{\cF}{{\cal F}}
\newcommand{\cX}{{\cal X}}
\newcommand\cg{\mathfrak{g}}
\newcommand{\IA}{\mathbb{A}}
\newcommand{\IP}{\mathbb{P}}
\newcommand{\IQ}{\mathbb{Q}}
\newcommand{\IH}{\mathbb{H}}
\newcommand{\IR}{\mathbb{R}}
\newcommand{\IC}{\mathbb{C}}
\newcommand{\IF}{\mathbb{F}}
\newcommand{\IV}{\mathbb{V}}
\newcommand{\II}{\mathbb{I}}
\newcommand{\IZ}{\mathbb{Z}}
\newcommand{\re}{{\rm Re}}
\newcommand{\im}{{\rm Im}}
\newcommand{\li}{{\rm Li}}
\newcommand{\Null}{{\rm null}}

\newcommand{\CA}{\mathbb A}
\newcommand{\CP}{\mathbb P}
\newcommand{\tmat}[1]{{\tiny \left(\begin{matrix} #1 \end{matrix}\right)}}
\newcommand{\mat}[1]{\left(\begin{matrix} #1 \end{matrix}\right)}
\newcommand{\diff}[2]{\frac{\partial #1}{\partial #2}}
\newcommand{\gen}[1]{\langle #1 \rangle}

\newcommand{\drawsquare}[2]{\hbox{%
\rule{#2pt}{#1pt}\hskip-#2pt
\rule{#1pt}{#2pt}\hskip-#1pt
\rule[#1pt]{#1pt}{#2pt}}\rule[#1pt]{#2pt}{#2pt}\hskip-#2pt
\rule{#2pt}{#1pt}}
\newcommand{\fund}{\raisebox{-.5pt}{\drawsquare{6.5}{0.4}}}
\newcommand{\antifund}{\overline{\fund}}

\newtheorem{theorem}{\bf THEOREM}
\def\thetheorem{\thesection.\arabic{theorem}}
\newtheorem{proposition}{\bf PROPOSITION}
\def\thetheorem{\thesection.\arabic{proposition}}
\newtheorem{observation}{\bf OBSERVATION}
\def\thetheorem{\thesection.\arabic{observation}}
\newtheorem{example}{\bf EXAMPLE}
\def\thetheorem{\thesection.\arabic{example}}

\def\theequation{\thesection.\arabic{equation}}
\newcommand{\setall}{\setcounter{equation}{0}
        \setcounter{theorem}{0}}
\newcommand{\setequation}{\setcounter{equation}{0}}
\renewcommand{\thefootnote}{\fnsymbol{footnote}}

\newcommand\Var{{\mathcal V}}
\newcommand\sP{{\mathcal P}}
\newcommand\sG{{\mathcal G}}
\newcommand\sL{{\mathcal L}}
\newcommand\sH{{\mathcal H}}

~\\
\vskip 1cm

\centerline{{\LARGE \bf Numerical Analyses on}}
\vskip 0.5cm
\centerline{{\LARGE \bf Moduli Space of Vacua}}
\medskip

\vspace{.4cm}

\centerline{
{\large Jonathan Hauenstein}$^1$,
{\large Yang-Hui He}$^2$ \&
{\large Dhagash Mehta}$^3$
}
\vspace*{3.0ex}

\begin{center}
{\it
{\small
{${}^{1}$
Department of Mathematics, North Carolina State University, \\
Raleigh, NC 27695-8205, USA \\
\qquad hauenstein@ncsu.edu\\
}
\vspace*{1.5ex}
{${}^{2}$
School of Physics, NanKai University, Tianjin, 300071, P.R.~China;\\
Department of Mathematics, City University, London, EC1V 0HB, UK;\\
Merton College, University of Oxford, OX14JD, UK\\
\qquad hey@maths.ox.ac.uk\\
}
\vspace*{1.5ex}
{${}^{3}$
Department of Physics, Syracuse University, \\
Syracuse, NY 13244, USA.\\
\qquad dbmehta@syr.edu
}
}}
\end{center}

\vspace*{4.0ex}
\centerline{\textbf{Abstract}}
\noindent We propose a new computational method to understand the vacuum moduli space of (supersymmetric) field theories.
By combining numerical algebraic geometry (NAG) and elimination theory, we develop a
powerful, efficient, and parallelizable algorithm to extract important information such as
the dimension, branch structure, Hilbert series and subsequent operator counting, as well as variation according to coupling constants and mass parameters.
We illustrate this method on a host of examples from gauge theory, string theory, and algebraic geometry.

\newpage

\tableofcontents


\section{Introduction}

The vacuum is undoubtedly one of the key ingredients in any quantum field theory, therefrom arises particles and dynamics.
Vacua which have non-trivial geometry are familiar to us, e.g., the famous ``Mexican hat'' of the Higgs potential
has its true minimum at a circle rather than a mere point (a local minimum at the origin).
In general, the expectation values of scalar fields can parameterize intricate manifolds called the vacuum moduli space.
In field theories with supersymmetry, where such scalar fields abound, the situation become particularly pronounced
and complicated geometries can arise as the vacuum moduli space.
In the context of string theory, realizations and interpretations of the geometry of the vacuum as the low-energy
limit of a compactification or holographic scenario is key to the engineering of field theories.
Indeed, one could even begin to unravel unexpected structure in the standard model by examining the vacuum under this light \cite{Gray:2005sr,Gray:2006jb,Hanany:2010vu}.
Thus, to have a mathematical tool to study the geometry, ranging from topological to metrical issues,
of the moduli space of vacua of field theories is therefore much needed.

Computational algebraic geometry has become one of the most useful tools to study numerous phenomena in theoretical physics.
In the recent years, active research has been concentrated on the rich interplay between algorithmic geometry and theoretical physics, especially in gauge and string theory \cite{comp-book}.
With the increasingly powerful computers the methods and algorithms to use algebraic geometry in practice to solve problems arising from theoretical physics have been becoming more relevant than ever.
Harnessing these developments to study the moduli space of vacua has been initiated over the past few years \cite{Gray:2006jb,Gray:2008yu}.

The traditional computational methods are based on symbolic computational algebraic geometry, most of whose sub-methods and sub-algorithms rely on
Gr\"obner basis techniques (cf.~\cite{Gray:2009fy} for a nice introduction for physicists).
Roughly speaking, given a vanishing set of the polynomials,
the so-called Buchberger Algorithm (BA) or its refined variants
compute a new equivalent system of polynomials, called a Gr\"obner basis \cite{CLO:07}, which has nicer properties; this is analogous to Gaussian elimination for linear systems.
Nowadays, efficient variants of the BA are available, e.g., F4 \cite{Faugere99anew}, F5 \cite{Faug:02}, and Involution Algorithms~\cite{2005math1111G}.
Symbolic computation packages such as {\sf Mathematica}, {\sf Maple}, {\sf Reduce}, etc., have built-in commands to calculate a Gr\"obner basis.
Moreover, {\sf Singular}
\cite{DGPS}, {\sf COCOA} \cite{CocoaSystem}, and {\sf Macaulay2} \cite{M2}
are specialized packages for computational algebraic geometry,
available as freeware, and {\sf MAGMA} \cite{BCP:97} is also such a specialized package available commercially.

In \cite{Gray:2006gn,Gray:2008zs,Gray:2007yq},
the Gr\"obner basis method was used to answering
questions pertinent to string and particle phenomenology and a publicly available computational package, called {\sf Stringvacua}, was designed to interface {\sf Mathematica} with {\sf Singular} for convenient usage.
Utilizing {\sf Stringvacua},  extracting such important information as the dimension of the vacuum, the number of real roots in the system, stability and supersymmetry of the potential, or the branches of
moduli space of vacua, etc.~can be readily settled using only a regular desktop machine in many circumstances.

Nonetheless, there are a few well-known problems with such methods.
First, the BA is known to suffer from {\it exponential complexity}: the computation time and the RAM required by the machine increases exponentially with the number of variables, equations, degree, and terms in each polynomial.
Moreover, BA is usually less efficient for systems with irrational coefficients.
One habitually has to resort to randomizing over the space of integer (or prime) coefficients and work over finite fields.  Another shortcoming is the highly sequential nature of the BA.

Although Gr\"obner methods have been an immensely useful tool to study various theoretical physics problems,
more prominently so in string and gauge theories, the time has come when one needs to depart from the computation of relatively simple models
and develop approaches for attacking more elaborate systems.
Unfortunately, methods based on BA often run out of the steam for more realistic models due to the aforementioned shortcomings.

A recently developed approach called numerical algebraic geometry (NAG) overcomes many of the shortcomings of Gr\"obner basis techniques.
The core method in NAG is numerical polynomial homotopy continuation (NPHC): given a system of polynomial equations which is known,
one computes its isolated solutions by first solving a related system and then tracking ``solution paths''.
More specifically, one first estimates an upper bound on the number of isolated solutions of the given system, creates a new system which is easy
to solve and has the same number of isolated solutions as this estimated upper bound, and finally one tracks the paths from each of the solutions of the
new system to the original system.  The paths which converge yield solutions of the original system.  In this way, for a system
which is known to have only finitely many solutions, one can obtain \textit{all} its solutions.
This is a remarkable method in its own right based on this guarantee to find all of the solutions
unlike other numerical methods such as the Newton-Raphson or its sophisticated variants.
Unlike standard Gr\"obner basis techniques, the NPHC method is \textit{embarrassingly parallelizeable}
and hence one can often solve more complicated systems efficiently using computer clusters.

In many theoretical physics problems, one encounters systems of equations that have infinitely many solutions.
In this case, the problem of solving such a system is transformed to a collection of systems of
equations which have isolated solutions corresponding to the components of the original system
and then invoke the NPHC method.
There are several sophisticated numerical packages well-equipped with
path trackers such as {\sf Bertini} \cite{BHSW06}, {\sf PHCpack}~\cite{Ver:99},
{\sf PHoM}~\cite{GKKTFM:04}, {\sf HOMPACK}~\cite{MSW:89}, and {\sf HOM4PS2}~\cite{GLW:05,Li:03},
which are all available as freewares from the respective research groups.

NAG was introduced in particle theory and statistical mechanics areas in Ref.~\cite{Mehta:2009}.
Subsequently, the NPHC method was used to solve systems arising in numerous physical phenomena
in lattice field theories \cite{Mehta:2009zv,Hughes:2012hg},
statistical physics \cite{Mehta:2011xs,Kastner:2011zz,PhysRevE.85.061103,Nerattini:2012pi},
particle phenomenology \cite{Maniatis:2012ex,CamargoMolina:2012hv}, and string phenomenology \cite{Mehta:2011wj,Mehta:2012wk,MartinezPedrera:2012rs,He:2013yk}.

Recently, we undertook the systematic introduction of  efficient methods in NAG to the
investigation of classes of problems in gauge and string theories \cite{Mehta:2012wk}.
There we focused particularly on systems of equations with finitely many solutions as these
are important to phenomenological questions such as looking for isolated extrema of potentials.
In some sense, what we undertake in our present study of moduli space of vacua is the substantial
generalization thereof where we can proceed from zero-dimensional solution sets to solution sets of arbitrary dimension.
We show that studying moduli spaces from an algebraic geometric perspective reduces to a problem in elimination theory
and then utilize the power of NAG to address this problem.

In particular, we propose a complete algorithm in this paper for computing the irreducible decomposition
of an elimination ideal, which is based on recently developed techniques regarding projections in
NAG \cite{Projection,ProjectMembership}.  We also combine classical methods in algebraic geometry
with NAG to compute the Hilbert function and Hilbert series of components of the solution set
which satisfy a certain regularity condition that holds for a wide range of problems in quantum field theory,
namely the components are arithmetically Cohen-Macaulay (see \cite[Chap. 1]{Migliore} for more details).
This is a mild technical constraint which most algebraic varieties we encounter in physics obey.

The outline of the paper is as follows.
We begin, in Section \ref{s:SVMS}, with a pedagogical introduction of
vacuum moduli spaces in supersymmetric field theories and explain why elimination theory of ideals is the most efficient approach.
Next, we turn to the mathematical problem of numerical elimination theory and propose a novel algorithm to compute the irreducible decomposition of elimination ideals in Section \ref{s:NAG}.
In Section \ref{s:eg}, we return to the physics and attack various explicit problems using our algorithm, ranging from quiver gauge theories to Calabi-Yau manifolds and from string phenomenology
to instantons.  These demonstrate that our numerical elimination algorithm is a powerful approach by solving problems well beyond the reach of traditional~methods.

\section{Supersymmetric Vacuum Moduli Space}\label{s:SVMS}\setall
As emphasized in the introduction, the study of the moduli space of vacua for gauge theories, especially ones of supersymmetry in (3+1)-dimensions, is of crucial importance to such diverse disciplines as string theory, particle phenomenology and mathematics.
In this section, we will therefore turn to a fairly pedagogical presentation of computing the vacuum moduli space, reflecting the three paradigmatic shifts in approach: the original problem (cf.~\cite{wess1992supersymmetry}), the symplectic quotient method of Luty-Taylor \cite{Luty:1995sd}, and the algebro-geometric algorithm of \cite{Gray:2005sr,Gray:2006jb}.

Our starting point is the action for an $\cN=1$ globally supersymmetric gauge theory in $3+1$-dimensions, written as an integral in superspace (with repeated indices implicitly summed):
\begin{equation}
\label{action}
S = \int d^4x\ \left[ \int d^4\theta\ \Phi_i^\dagger e^V \Phi_i +
    \left( \frac{1}{4g^2} \int d^2\theta\ \tr{W_\alpha W^\alpha} +
    \int d^2\theta\ W(\Phi) + {\rm h.c.} \right) \right]
.
\end{equation}
Here, the $\Phi_i$ are chiral superfields transforming in some representation $R_i$ of gauge group $G$; $V$ is a vector superfield transforming in the Lie algebra $\cg$ of $G$; $W_\alpha = i\overline{D}^2 e^{-V} D_\alpha e^V$, the gauge field strength, is a chiral spinor superfield; and $W(\Phi)$ is the superpotential, which is a holomorphic function of the $\Phi_i$.

The supersymmetric vacuum moduli space (VMS) $\cM$ is simply the set of field configurations which minimize the effective potential obtained from integrating out the Grassman variables $\theta_i$ in \eqref{action}.
Explicitly \cite{argyres}, they are the solutions to
\begin{equation}\label{FD}
\left\{
\begin{array}{lll}
\mbox{F-Flatness} &:& \cF = \{ \frac{\partial}{\partial \phi_i} W(\phi_i) = 0  \} ; \\
\mbox{D-Flatnes}  &:& D^A = \sum\limits_i \phi_i^\dagger T^A \phi_i = 0 \ .
\end{array}
\right.
\end{equation}
with $\phi_i$ being the vacuum expectation values of the scalar components of $\Phi_i$.
In the D-flatness conditions, we have chosen the Wess-Zumino gauge and $T^A$ are the generators of
$\cg$.
The index $i$ runs over the fields and $A$, over the gauge group generators.

There is gauge redundancy to the action in \eqref{action}, namely
$\Phi \rightarrow \exp(i \Lambda) \cdot \Phi; \ e^V \rightarrow \exp(- i \Lambda^\dagger)
\cdot  e^V \cdot \exp(- i \Lambda)$, in terms of the parameter $\Lambda \in \cg$.
Choosing $\Lambda$ to be complex makes the full gauge symmetry to be the complexification
$G^c$ of the original gauge group $G$.
Following a series of systematic studies of fixing this gauge symmetry \cite{Buccella:1982nx,Gatto:1986bt,Procesi:1985hr,Witten:1993yc}, it was realized in \cite{Luty:1995sd}
that for every solution to the F-flatness constraints there is one and only one solution to the
D-flatness conditions.
Therefore, D-flatness is merely a gauge-fixing condition and the full set of solutions to \eqref{FD}
is simply a symplectic quotient of the space of F-flatness by the complexified group $G^c$.
In other words, as an algebraic variety, the VMS is the GIT quotient
\begin{equation}
\cM = \cF ~//~ G^c .
\end{equation}

From a computational algebraic geometric point of view, this rather abstract quotient was made explicit in \cite{Gray:2006jb}.
Let there be $n$ superfields $\Phi_{i=1,\ldots,n}$ with associated vacuum expectation value
of scalar component $\phi_i$ and polynomial superpotential $W(\phi_i)$.
This allows us to define the F-flatness equations as the Jacobian ideal $\gen{\frac{\partial}{\partial \phi_i}{W}}$ in the polynomial ring $R = \IC[\phi_1, \ldots, \phi_n]$.

Since the D-flatness equations are gauge fixing conditions, we can consider the set of gauge invariants, composed of polynomials in $\phi_i$.
These are traces of matrix products so that the final answer will carry no free gauge index.
Moreover, there should be a minimal generating set of the ring of such invariants, which we denote as
$D = \{ r_j ( \{ \phi_i \}) \}$ where $r_{j=1, \ldots,k}$ are polynomials and $k$ is a positive integer (potential quite large).
Hence, we have another polynomial ring $S = \IC[r_1, \ldots, r_k]$.

The realization in \cite{Gray:2006jb} is that the polynomial map $D$ is a ring map from the quotient ring $\cF \simeq R \left/ \gen{\frac{\partial}{\partial \phi_i}{W}} \right.$ to the ring $S$:
\begin{equation}
\IC[\phi_1, \ldots, \phi_n] \left/ \left\langle\frac{\partial}{\partial \phi_i}{W}
\right\rangle\right. \qquad
\stackrel{D = \{ r_j ( \{ \phi_i \}) \}}{\xrightarrow{\hspace*{3cm}}}
\qquad
\IC[r_1, \ldots, r_k]  \ .
\end{equation}
The image of this ring map is then the VMS, defined as an affine variety in $S$:
\begin{equation}\label{Dmap}
\cM \simeq \im(\cF \stackrel{D}{\longrightarrow} S) \ .
\end{equation}

The above procedure is perfectly adapted for the language of computational
commutative algebra, especially using either of the wonderful computer packages
{\sf Macaulay 2} \cite{M2} and {\sf Singular} \cite{GBBIB726}.
However, the Gr\"obner basis algorithm is central to this computation
and is precisely the computationally intensive step we wish to circumvent.
To this end, one can use a standard trick in polynomial manipulations to re-phrase and simplify the above process (cf. \S 5 of \cite{Gray:2009fy}).
Instead of considering $D$ as a map from $R$ to $S$, we can simply enlarge the ring $R$ by adjoining new variables $y_{j = 1, \ldots, k}$ to form the larger ring
$\tilde{R}$ and consider the following ideal therein:
\begin{equation}
\left\langle
\frac{\partial {W}}{\partial \phi_i} \ ,
y_j - r_j( \{ \phi_i \})
\right\rangle
\subset
\tilde{R} = \IC[\phi_{i=1, \ldots, n}, y_{j = 1, \ldots, k}] \ .
\end{equation}

The image in \eqref{Dmap} describes the relations (first syzygy) amongst the variables $r_j$ subject to the F-flatness conditions of $\cF$.
This is equivalent to systematic elimination of the $\phi_i$ variables in $\tilde{R}$ so that only the $y_j$ variables remain.
The resulting ideal in $\tilde{R}$ will explicitly have only polynomial relations amongst the $y_j$ variables.
Since we have set each $y_j$ equal to $r_j(\{\phi_i\})$, these relations define precisely the VMS in $\tilde{R}$.
Hence, this formulation bypasses the language of ring maps entirely and we are now confronted with an {\bf elimination problem}.

Indeed, there are classical and modern techniques for elimination,
such as using resultants and computing the Gr\"obner basis of the ideal with a so-called
{\em elimination ordering} for the variables.  Although it is impossible for us to provide
a complete account of techniques in elimination theory, we will provide a little historical perspective.
Gaussian elimination is the standard elimination technique for systems of linear equations.
In simple terms, Buchberger's algorithm can be thought of as a generalization of Gaussian elimination for polynomial systems.
As mentioned above, resultants were classically applied to perform eliminations and there are many renowned
mathematicians associated with them including B\'ezout, Euler, Sylvester, Cayley, Dixon, and Macaulay,
and reinvigorated by Lazard.  Another recent approach that is related to, but independent of, Gr\"obner bases
is Wu's characteristic sets.  We refer the reader to \cite{EliminationMethods} and the references therein
for more details on these methods and related history.

In this paper, continuing with the experience gained from \cite{Mehta:2012wk}, we will try to avoid the
expensive step of computing Gr\"obner bases when all the information needed for the VMS are such quantities as
the dimension, irreducible components, and even Hilbert series.
Instead, we will harness the power of numerical algebraic geometry
in \sref{s:NAG} to perform this computation.

\subsection{Summary of Algorithm}

It is expedient to summarize the algorithm which we employ to compute the~VMS:

\fbox{
\begin{varwidth}{20cm}
\begin{itemize}
\item {\bf INPUT:}
\begin{enumerate}
\item Superpotential $W(\{\phi_i\})$, a polynomial in variables $\phi_{i=1, \ldots, n}$,
\item Generators of gauge invariants: $r_j(\phi_i)$, $j=1, \ldots, k$ polynomials in $\phi_i$;
\end{enumerate}

\item {\bf ALGORITHM:}
\begin{enumerate}
\item Define the polynomial ring $\tilde{R} = \IC[\phi_{i=1,\ldots,n}, y_{j=1,\ldots,k}]$,
\item Consider the ideal $I = \gen{\frac{\partial {W}}{\partial \phi_i}; y_j - r_j(\phi_i)}$,
\item Eliminate all variables $\phi_i$ from $I \subset \tilde{R}$, giving the ideal $M$ in terms of $y_j$;
\end{enumerate}

\item {\bf OUTPUT:}
$M$ corresponds to the VMS as an affine variety in $\IC[y_1, \ldots, y_k]$.
\end{itemize}
\end{varwidth}
}

\section{Numerical Algebraic Geometry and Elimination}\label{s:NAG}
\setall
Having phrased our physics as a problem in mathematics, in this section, we will develop the necessary mathematical and computational tools.
The Gr\"obner basis approach presented in \sref{s:SVMS} uses techniques from algebra via manipulation
of equations to compute eliminations.  The numerical algebraic geometric techniques presented
here compute eliminations based on geometry via points.  The first two techniques
presented are for computing low degree polynomials, which can be used to improve both Gr\"obner basis
and additional numerical algebraic geometric computations.  The other two techniques utilize
the numerical algebraic geometric notion of witness sets (see \cite[Chap. 13]{SW:05}) to compute the elimination.
We close the section by computing the Hilbert series in the
arithmetically Cohen-Macaulay case (for more details, see \cite[Chap. 1]{Migliore}).
An ongoing project is a practical method for computing the Hilbert series in general.

For a polynomial system $F$, define $\Var(F) = \{x~|~F(x) = 0\}$ to be the variety defined by $F$.
To simplify the presentation, throughout this section, we
let $\pi(x,y) = y$ and $Z = \Var(\sG)$ for some polynomial system $\sG$.
We aim to compute the elimination $Y = \overline{\pi(Z)}$ which
will be considered based on the following three cases:
\begin{enumerate}
\item {\bf Parametrization:} $\sG(x,y) = y - F(x)$;
\item {\bf Restricted parameterizations:} $\sG(x,y) = \left[\begin{array}{c} G(x) \\ y - F(x) \end{array}\right]$; and
\item {\bf General case:} $\sG(x,y)$ is some polynomial system,
\end{enumerate}
where $F$ and $G$ are polynomial systems.  In the parameterized case, $Z$ is an irreducible and smooth variety
implying that $Y$ is also irreducible.  In the other two cases, irreducibility
depends upon the structure of the polynomial system $\sG$.  Additionally, $\sG$ may
impose multiplicities on the irreducible components of $Z$, e.g., $\sG(x,y) = (x-y)^2$.
However, by using isosingular deflation \cite{Isosingular}
where appropriate, we can assume without loss of generality that each irreducible component of $Z$ has
multiplicity one with respect to $\sG$.
This allows us to compute dimensions simply by computing the dimension of a linear space related to the Jacobian of $\sG$
as shown in the following section.

\subsection{Dimension of the image}\label{s:Dim}

Let $V\subset Z = \Var(\sG)$ be an irreducible component of multiplicity one with respect to $\sG$
and $(\hat{x},\hat{y})\in\IC^N\times\IC^M$ be a general point of $V$.  Then, the dimension of the irreducible algebraic set $\overline{\pi(V)}$
(which may or may not be an irreducible component of $Y = \overline{\pi(Z)}$) is easily computed via linear algebra \cite{Projection} as follows:
\begin{equation}\label{eq:Dim}
\dim \overline{\pi(V)} = \dim \Null J\sG(\hat{x},\hat{y}) - \dim \Null J\sG(\hat{x},\hat{y})_{[\cdots,1:N]}
\end{equation}
where $J\sG(\hat{x},\hat{y})$ is the Jacobian matrix of $\sG$ at $(\hat{x},\hat{y})$ and $J\sG(\hat{x},\hat{y})_{[\cdots,1:N]}$ is the matrix
consisting of the first $N$ columns of $J\sG(\hat{x},\hat{y})$.
In the parameterized case, namely $\sG(x,y) = y - F(x)$, we know that $Z$ and $Y$ are both
irreducible with $\dim Z = N$ and
\begin{equation}\label{eq:DimSpecial}
\dim Y = N - \dim \Null JF(\hat{x}).
\end{equation}

\begin{example}\label{ex:Test}
We consider the parameterized example $\sG(x,y) = y - F(x)$ where
$$F(x_1,x_2,x_3,x_4,x_5,x_6,x_7,x_8) = \left[\begin{array}{c}
x_2x_3-x_1x_4 \\
x_1x_5+x_2x_6 \\
x_3x_5+x_4x_6 \\
x_1x_7+x_2x_8 \\
x_3x_7+x_4x_8 \\
x_6x_7-x_5x_8 \end{array}\right].$$
The set $Z = \Var(\sG)$ is irreducible of dimension $8$.  The dimension
of $Y = \overline{\pi(Z)}$ is $8-3=5$ since $\dim \Null JF(\hat{x}) = 3$
for a random $\hat{x}\in\IC^8$.
\end{example}

\subsection{Low degree polynomials via interpolation}\label{s:Interp}

In order to use interpolation techniques to compute low degree polynomials, one must have the ability to compute points on $Y$.
In the parameterized case, computing points on $Y$ is trivial since $F(x)\in Y$ for every $x$.
Under the assumption that $Z$ is irreducible, the numerical algebraic geometric technique of sampling an
irreducible component \cite[\S 15.2]{SW:05} allows one to compute arbitrarily many points on $Y$
given a sufficiently general point $(x,y)\in Z$.
If $Z$ is not irreducible, then we need a sufficiently general point on each irreducible component of $Z$.
In short, sampling $Z$, and hence $Y$ via the projection $\pi$, reduces to following a curve
starting at $(x,y)$ defined by the intersection of $Z$ with a family of linear spaces
of complimentary dimension to $Z$.  Additionally, we restrict our attention to low degree polynomials
due to the exponential growth of the matrices involved and to avoid issues regarding numerical stability.
Nonetheless, computing some low degree polynomials can be used to expedite further computations.

Given a finite dimensional linear space of polynomials $\sP$,
interpolation using sufficiently many points on each of the irreducible components of $Y$
yields the subspace of polynomials in $\sP$ which vanish on $Y$.
This is accomplished by computing the null space of a matrix
constructed by evaluating the points at a basis for $\sP$.  We note that using more points,
thus adding additional rows to this matrix, and rescaling each row
will often drastically improve the conditioning of this matrix, e.g., see \cite{HilbertZero}.
Additionally, once such a polynomial is computed, it is often easy to use either symbolic evaluation
or Gr\"obner basis methods to validate that a given polynomial does indeed vanish~on~$Y$.

\begin{example}\label{ex:TestInterp}
For the elimination problem in Ex.~\ref{ex:Test}, the set $Y$ was verified to be
a hypersurface in $\IC^6$ meaning that $Y = \Var(f)$ for some polynomial $f$.
Before we consider using interpolation to compute this polynomial $f$,
we note that since each polynomial in $F$ is homogeneous of degree $2$, $f$ is also homogeneous.
Thus, we will restrict our attention to homogeneous polynomials in $y_1,\dots,y_6$.

We first consider $\sP$ to be the set of linear homogeneous polynomials in $y_1,\dots,y_6$
taking the variables as the basis.
Each row of the matrix under consideration is thus of the form
\begin{equation}\label{eq:Linear}
\left[\begin{array}{cccccc} F_1(\hat x) & F_2(\hat x) & F_3(\hat x) & F_4(\hat x) & F_5(\hat x) & F_6(\hat x) \end{array}\right]
\end{equation}
where $\hat x \in \IC^8$ since $y_i = F_i(x)$.  After picking 6 random points in $\IC^8$, it is easy to verify that
the resulting $6\times6$ matrix has full rank meaning that there are no linear polynomials
which vanish on $Y$.

We then consider $\sP$ to be the set of homogeneous polynomials in $y_1,\dots,y_6$ of degree $2$ taking the monomials of degree $2$,
namely $y_1^2, y_1y_2, y_1y_3, \dots, y_6^2$, as the basis.
Each row of the matrix under consider is thus of the form
\begin{equation}\label{eq:Quadratic}
\left[\begin{array}{ccccc} F_1(\hat x)^2 & F_1(\hat x) F_2(\hat x) & F_1(\hat x) F_3(\hat x) & \cdots & F_6(\hat x)^2 \end{array}\right].
\end{equation}
where $\hat x\in\IC^8$.
After picking 21 random point in $\IC^8$, it is easy to verify that
the resulting $21\times21$ matrix has rank $20$ meaning that there is a quadratic polynomial
which vanishes on $Y$.  Computing the null space reveals this polynomial to be
$y_1y_6 - y_2y_5 + y_3y_4$.  By substitution, it is easy to verify that
$F_1 F_6 - F_2 F_5 + F_3 F_4 = 0$.
\end{example}

\subsection{Low degree polynomials via lattice base reduction}\label{s:LLL}

If the polynomials defining $Z$ have rational coefficients,
then the same is true for $Y$.  The immediately follows
from the fact that methods involving only Gr\"obner bases maintain
the same field of coefficients.  As described in \cite[\S~3.3]{RecoveryNumSym},
lattice base reduction techniques, such as LLL \cite{LLL} and PSLQ \cite{PSLQ},
provide an approach to compute polynomials vanishing on $Y$.
For this technique, only one sufficiently general point is needed
for each irreducible component of $Y$.

Fix a sufficiently general point $y$ on the irreducible component $V\subset Y$.
As in \sref{s:Interp}, we want to compute the subspace of polynomials
in a finite dimensional linear space of polynomials $\sP$ which vanish on $V$.
Let $v$ be the vector obtained by evaluating $y$ on a basis for $\sP$.
Lattice base reduction techniques compute an integer vector $w$ such that $v\cdot w\approx 0$.
By changing the precision and recomputing, one is able to reliably determine if
$w$ actually corresponds to a polynomial which vanishes on $V$.
Moreover, as stated above, validation is often trivial.

\begin{example}\label{ex:Test2}
Consider repeating the computation from Ex.~\ref{ex:Test} using lattice base reduction
rather than numerical interpolation.  We proceed by first fixing a random $\hat x\in\IC^8$.
For the linear polynomials, we apply PSLQ to the vector computed in \eqref{eq:Linear}.
Using 20 digits, this yields $w = [-431, -156, 571, 300, 597, 95]$
which changes to $w = [8994, 16067, -12869, -33044, 27787, 56416]$ when using 30 digits.
This suggests that there are no linear relations with small integer coefficients.

For the quadratic polynomials, PSLQ applied to the vector computed in \eqref{eq:Quadratic},
using both 20 and 30 digits, computes the vector $w$ corresponding to $y_1y_6 - y_2y_5 + y_3y_4$.
\end{example}

\subsection{Solve by slicing}\label{s:Slicing}

Suppose that we are in the restricted parameterizations case, which includes the parameterizations
case by simply taking $G(x)$ to be the zero polynomial.  We assume $G:\IC^N\rightarrow\IC^n$
and $F:\IC^N\rightarrow\IC^M$ are polynomial systems and $V\subset\Var(G)$ is an irreducible component
(where $V = \IC^N$ in the parameterized case).
By using isosingular deflation \cite{Isosingular} if needed, we can assume that $V$ has multiplicity one with respect to $G$.
The key numerical data structure for $V$ is a {\em witness set} which is the triple
$\{G,L,W\}$ where $L:\IC^N\rightarrow\IC^\ell$ is a system of general linear polynomials where $\ell = \dim V$
and $W = V\cap\Var(L)$.  The set $W$, called a {\em witness point set}, consists of $\deg V$~points.

The set $\overline{F(V)}$ is irreducible since $V$ is irreducible and $F$ is polynomial.
Its dimension can be computed using \eqref{eq:Dim}, say $\ell' = \dim \overline{F(V)}$.
Since we do not have a system of polynomials for which $\overline{F(V)}$ is an irreducible component,
we can not readily compute a witness set for $\overline{F(V)}$.  However,
we are able to compute a {\em pseudo-witness set} \cite{Projection} for $\overline{F(V)}$
which allows us to answer questions about $\overline{F(V)}$, such as its degree.
We do this by considering the following polynomial system constructed by adding slices to~$\sG$:
\begin{equation}\label{eq:SliceSys}
\sH(x,y) = \left[\begin{array}{c} G(x) \\ \sL(x) \\ y - F(x) \\ \widehat\sL(y) \end{array}\right]
\end{equation}
where $\sL:\IC^N\rightarrow\IC^{\ell-\ell'}$ and $\widehat\sL:\IC^M\rightarrow\IC^{\ell'}$ are systems
of general linear polynomials.  Let $\widehat W$ be the set consisting of the finitely
many solutions of $\sH$ whose $x$ coordinates lie on $V$.  Such a set may be computed
efficiently from a witness set for $V$ using regeneration \cite{HSW10}.
The quadruple $\{\sG,\pi,M,\widehat W\}$ where $M(x,y) = \left[\begin{array}{c} \sL(x) \\ \widehat\sL(y) \end{array}\right]$
is a pseudo-witness set for $\overline{F(V)}$ with
$\deg \overline{F(V)} = |\pi(\widehat W)|$.
For any point $y\in\overline{F(V)}$, the fiber over $y$ in $V$
is the set
$$F_y = \{x\in V~|~y = F(x)\}.$$
For a general point $y\in\overline{F(V)}$,
$\dim F_y = \ell - \ell'$ and $\deg F_y = |\widehat W|/|\pi(\widehat W)|$.

\begin{example}\label{ex:TestInterp2}
For the elimination problem in Ex.~\ref{ex:Test}, the set $Y = \overline{F(\IC^8)} = \overline{\pi(Z)}$
was verified to be a hypersurface in $\IC^6$ so that $\ell = 8$, $\ell' = 5$, and $\ell - \ell' = 3$.  To compute the
degree of $Y$ and the degree of the general fiber, we solve $\sH(x,y) = 0$ defined in \eqref{eq:SliceSys}
where we take $G(x)$ to the zero polynomial.  That is, $\sH$ is a polynomial system depending on $14$ variables
that consists of $3$ linear slices $\sL$, $6$ polynomials $y - F(x)$, and $5$ linear slices $\widehat\sL$.
The set $\widehat W$ consists of $4$ points with $|\pi(\widehat W)| = 2$.
Therefore, $\deg Y = 2$ and the general fiber has dimension $3$ and degree $4/2 = 2$.
\end{example}

To compute the irreducible components of $Y$, we first compute a
witness set for each irreducible component of $\Var(G)$, called a {\em numerical irreducible decomposition}
(see \cite[Chap. 15]{SW:05} for more details).
Then, for each irreducible component $V\subset\Var(G)$, we
follow the aforementioned approach to compute a pseudo-witness set for $\overline{F(V)}$.
The irreducible components of $Y$ are precisely the inclusion maximal sets of
$$\left.\left\{\overline{F(V)}~\right|~V\subset\Var(G)\hbox{~is an irreducible component}\right\}.$$
These inclusion maximal sets can be computed using the pseudo-witness sets constructed above
with the membership test provided in \cite{ProjectMembership}.

\subsection{General approach}\label{s:GeneralApproach}

The general case follows a similar approach as above except that we must
start with a numerical irreducible decomposition of $Z = \Var(\sG)$, that is,
compute a witness set for each irreducible component of $Z$.
For each irreducible component $V$, we follow the basic approach \cite{Projection}
to construct a pseudo-witness set for $\overline{\pi(V)}$.  Then,
the irreducible components of $Y = \overline{\pi(Z)}$ are precisely the inclusion maximal sets of
$$\left.\left\{\overline{\pi(V)}~\right|~V\subset\Var(\sG)\hbox{~is an irreducible component}\right\}$$
which can be computed using the membership test of \cite{ProjectMembership}.

\begin{example}\label{ex:TestInterp3}
For a basic example, consider computing the closure of the set of all $(a,c)\in\IC^2$
such that $ax^2+2ax+c$ has at least one root where the derivative with respect to $x$ vanishes.
That is, we want to compute $Y = \overline{\pi(Z)}$ where
$$\sG(x,a,c) = \left[\begin{array}{c} ax^2 + 2ax + c \\ 2ax + 2a \end{array}\right],~~Z = \Var(\sG), \hbox{~~and~~} \pi(x,a,c) = (a,c).$$
A numerical irreducible decomposition shows that $Z$ decomposes into a line and quadratic curve,
namely $V_1 = \{(x,0,0)~|~x\in\IC\}$ and $V_2 = \{(x,a,1)~|~ax=-1\}$, respectively.
Using \sref{s:Dim}, we have that $\dim \overline{\pi(V_1)} = 0$ and $\dim \overline{\pi(V_2)} = 1$.
Since $\overline{\pi(V_1)}$ is the singleton containing the origin,
a witness set for $V_1$ trivially yields a pseudo-witness set for $\overline{\pi(V_1)}$.
A pseudo-witness set for $\overline{\pi(V_2)}$ using \cite{Projection}, which simply moves linear slices,
yields that $\deg \overline{\pi(V_2)} = 1$ and the general fiber consists of one point.
In particular, $\overline{\pi(V_2)} = \{(a,1)~|~a\in\IC\}$.  The final step
uses \cite{ProjectMembership} to determine that $\overline{\pi(V_1)}$ is not contained in $\overline{\pi(V_2)}$
meaning $Y$ has two irreducible components, namely a point and~a~line.
\end{example}

\subsection{Hilbert series}\label{s:HilbertSeries}

Positive dimensional arithmetically Cohen-Macaulay (aCM) schemes retain
information under slicing by hyperplanes and hypersurfaces \cite[Chap. 1]{Migliore}.
In short, this implies that the Hilbert function and Hilbert series for an
aCM scheme can be computed directly from a witness point set.

We divert briefly into projective algebraic geometry.
Assume that $V\subset\IP^n$ is a closed subscheme of dimension $r\geq1$.
As above, we will assume that $V$ is generically reduced, i.e., simply consider
it as an algebraic set in $\IP^n$.  There are two equivalent ways that we will describe
that one can use to determine if $V$ is aCM.  The first, which is typically taken to be the definition of aCM,
is that the quotient $S/I_V$ is a Cohen-Macaulay ring where
$S$ is the homogeneous polynomial ring $\IC[x_0,\dots,x_n]$ and $I_V$ is the saturated ideal associated to $V$.
That is, the Krull dimension of $S/I_V$ is the same as the depth of $S/I_V$.
The second is that deficiency modules of $V$ are trivial,
namely $(M^i)(V) = H^i_*(\cI_V) = 0$ for $1 \leq i \leq r$ where $\cI_V$ is the ideal sheaf of $V$.

We denote the Hilbert function of $V$ using an integer
vector notation, namely $H_V = (h_V^0,h_V^1,\dots)$,
and the Hilbert series as the function $HS_V(t) = \sum_{j=0}^\infty h_V^j t^j$.

Let $W$ be a witness point set for $V$, that is, there exists a general codimension $r$ linear space $\sL$
such that $W = V\cap\sL$.  The Hilbert function for $W$, considered as simply a zero-dimensional set,
can be computed via numerical linear algebra using Veronese embeddings of the points of $W$, e.g., see \cite{HilbertZero}.
We know $h_W^0 = 1$, $h_W^j \geq 0$ for all $j\geq0$, and
$\sum_{j=0}^\infty h_W^j = |W|$ implying $h_W^j = 0$ for all $j$ sufficiently large.
In particular, the Hilbert series for $W$, namely $HS_W(t)$, is a polynomial.

The following theorem, which follows from Corollary 1.3.8(c) and Section 1.4 of \cite{Migliore},
 shows how to compute the Hilbert series and Hilbert function for $V$ from $H_W$.

\begin{theorem}
Let $V$ be a generically reduced and aCM scheme of dimension $r\geq0$.  Let $W$ be a witness point set for $V$
with Hilbert function $H_W = (h_W^0,h_W^1,\dots)$.  Then,
\begin{enumerate}
\item $H_V = (h_V^0,h_V^1,\dots)$ where $h_V^k = \displaystyle\sum_{j_1=0}^k \sum_{j_2=0}^{j_1} \cdots \sum_{j_r=0}^{j_{r-1}} h_W^{j_r}$; and
\item $HS_V(t) = \dfrac{HS_W(t)}{(1-t)^r}$.
\end{enumerate}
\end{theorem}

Even though we have stated this for witness point sets of projective varieties,
we can simply restrict to an affine patch by taking linear slices that are general with respect to the chosen patch.
In short, the same computation holds for affine varieties
and a generalization to pseudo-witness point sets is trivial.

\begin{example}\label{ex:TestHilbert}
Reconsider the elimination problem first considered in Ex.~\ref{ex:Test}.
We know that $Y$ is a hypersurface in $\IC^6$ of degree 2 that one can verify
is aCM.  Let $\sL_1,\dots,\sL_5$ be general hyperplanes in $\IC^6$ and
consider the witness point set $W = V\cap\sL_1\cap\cdots\cap\sL_5$ consisting of two points, say $W = \{w_1,w_2\}$.
Since the matrix $[w_1~w_2]$ has rank 2, we know $h_W^0 + h_W^1 = 2$ yielding $H_W = (1,1,0,0,\dots)$.
Table~\ref{tab:Hilbert} presents the Hilbert function and Hilbert series
for $V_k = V\cap\sL_1\cap\cdots\cap\sL_k$ for $k = 0,\dots,5$
which can be easily verified using either {\sf Macaulay 2} \cite{M2} or {\sf Singular} \cite{GBBIB726}.
In particular, $HS_V(t) = (1+t)/(1-t)^5$.

\begin{table}[!ht]
\centering
\begin{tabular}{|c|l|c|}
\hline
$k$ & \multicolumn{1}{|c|}{$H_{V_k}$} & $HS_{V_k}$ \\
\hline
$5$ & $(1,1,0,0,0,\dots)$ & $1+t$ \\
\hline
$4$ & $(1,2,2,2,2,\dots)$ & $\frac{1+t}{1-t}$ \\
\hline
$3$ & $(1,3,5,7,9,\dots)$ & $\frac{1+t}{(1-t)^2}$ \\
\hline
$2$ & $(1,4,9,16,25,\dots)$ & $\frac{1+t}{(1-t)^3}$ \\
\hline
$1$ & $(1,5,14,30,55,\dots)$ & $\frac{1+t}{(1-t)^4}$ \\
\hline
$0$ & $(1,6,20,50,105,\dots)$ & $\frac{1+t}{(1-t)^5}$ \\
\hline
\end{tabular}\caption{Summary of Hilbert functions and Hilbert series}\label{tab:Hilbert}
\end{table}
\end{example}

\section{Illustrative Examples and Applications}\label{s:eg}\setall
Having outlined the method of attack in the computation of the moduli space of vacua and discussed in detail the mathematical background underlying the algorithm, we now, in this section, turn to concrete examples which arise in actual physical situations and demonstrate how our numerical perspective affords an efficient outlook on understanding the physics.

\subsection{Warmup: an Orbifold Theory}
Let us start with a four-dimensional $\cN=1$ theory which is famous to the AdS/CFT literature.
This is the world-volume theory of a D3-brane transverse to the non-compact Calabi-Yau threefold locally realized as the quotient of $\IC^3$ by $\IZ_3$ which acts on the three complex coordinates $(x,y,z)$ of the former by $(x,y,z) \rightarrow \exp(\frac{2 \pi i }{3}) (x,y,z)$ (otherwise known as $(1,1,1)$ action).
The theory is given by a $U(1)^3$ quiver together with a superpotential $W$:
\begin{equation}
\begin{array}{cc}
\begin{array}{c} \includegraphics[trim= 0mm 0mm 0mm 150mm,clip,width=2.5in]{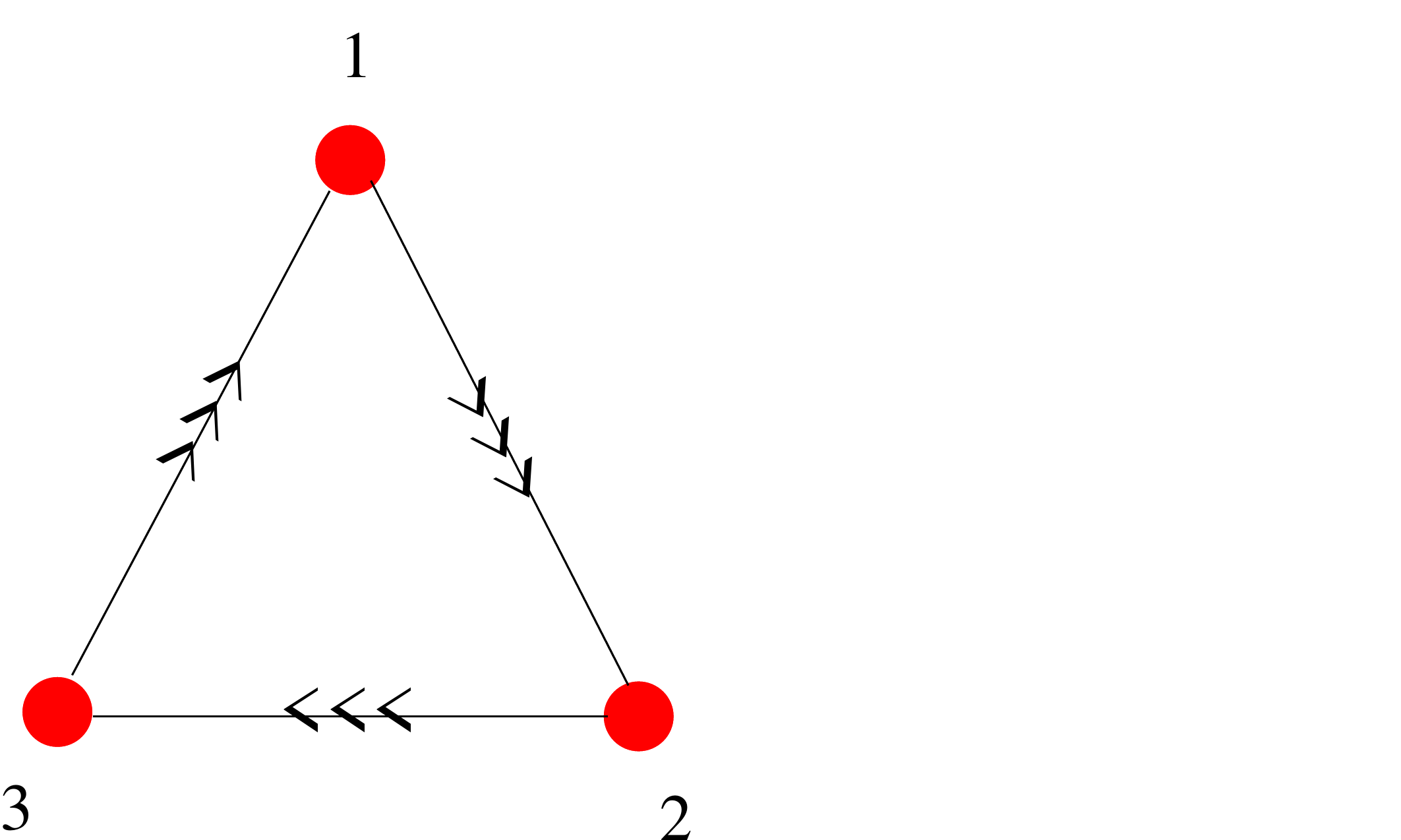} \end{array}
&
W= \sum\limits_{\alpha,\beta,\gamma=1}^3\epsilon_{\alpha\beta\gamma} X^{(\alpha)}_{12} X^{(\beta)}_{23} X^{(\gamma)}_{31}.
\end{array}
\end{equation}
There are nine fields $X^{(\alpha)}_{12}, X^{(\beta)}_{23}$, and $X^{(\gamma)}_{31}$, $\alpha,\beta,\gamma=1,2,3$, with the subscript $ij$ signifying an arrow from node $i$ to $j$ and superscript $\alpha$ denoting that there is a multiplicity of three arrows for each pair of nodes.
Subsequently, there are $3^3 = 27$ gauge invariant operators formed by their products corresponding to the closed cycles in the quiver.
In the superpotential, $\epsilon_{\alpha\beta\gamma}$ is the standard totally antisymmetric Levi-Civita symbol on three indices.
Taking the partial derivatives of $W$ with respect to the fields, we obtain 9 F-terms: $ \sum\limits_{\beta,\gamma=1}^3\epsilon_{\alpha\beta\gamma} X^{(\beta)}_{23}X^{(\gamma)}_{31}$,
$\sum\limits_{\alpha,\gamma=1}^3\epsilon_{\alpha\beta\gamma} X^{(\alpha)}_{12}X^{(\gamma)}_{31}$, and
$\sum\limits_{\alpha,\beta=1}^3\epsilon_{\alpha\beta\gamma} X^{(\alpha)}_{12}X^{(\beta)}_{23}$.

Of course the moduli space of vacua of F-flatness and D-flatness should give the affine equations for $\IC^3/\IZ_3$ as an algebraic variety; this was studied in detail in \S2.3.2 of \cite{Gray:2006jb}.
Here, in light of our new methodology, we are trying to solve the following problem (below, $\alpha, \beta, \gamma = 1,2,3$ with free indices ranging fully):
\begin{eqnarray}
\nn
\mbox{ Given ideal } &: &
\left\{
\begin{array}{l}
y_{\alpha\beta\gamma} - X^{(\alpha)}_{12} X^{(\beta)}_{23} X^{(\gamma)}_{31} \ , \\
\sum\limits_{\beta,\gamma=1}^3\epsilon_{\alpha\beta\gamma} X^{(\beta)}_{23}X^{(\gamma)}_{31}, \\
\sum\limits_{\alpha,\gamma=1}^3 \epsilon_{\alpha\beta\gamma} X^{(\alpha)}_{12}X^{(\gamma)}_{31}, \\
\sum\limits_{\alpha,\beta}^3\epsilon_{\alpha\beta\gamma} X^{(\alpha)}_{12}X^{(\beta)}_{23}\\
\end{array}
\right\} \subset \IC[y_{\alpha\beta\gamma}; X^{(*)}_{12}, X^{(*)}_{23}, X^{(*)}_{31}]
\\
\label{elim-dp0}
\mbox{ Eliminate } &: &
X^{(*)}_{12}, \ X^{(*)}_{23}, \ X^{(*)}_{31} \ .
\end{eqnarray}
This ideal consists of $3^3 + 3 \cdot 3 = 36$ generators in the polynomial ring with $3^3 + 9 = 36$ variables.
After elimination, we obtain relations amongst the 27 variables $y_{\alpha\beta\gamma}$.

Our first step is to compute the numerical irreducible decomposition of the $9$ generators involving
only the $X$ variables, which yields one irreducible component of dimension $5$ and degree $6$.
Therefore, the eliminant is irreducible and, using \eqref{eq:Dim}, has dimension $3$ with
the general fiber having dimension $2$.  Using the approach of \S~\ref{s:Interp}, one readily obtains
the following 17 linear and 27 quadratic equations:
{\footnotesize
$$\begin{array}{ccccccccccc}
 y_{112} - y_{211} &=& y_{113} - y_{311} &=& y_{121} - y_{211} &=& y_{122} - y_{221} &=& y_{123} - y_{321}  \\
                   &=& y_{131} - y_{311} &=& y_{132} - y_{321} &=& y_{133} - y_{331} &=& y_{212} - y_{221} \\
                   &=& y_{213} - y_{321} &=& y_{223} - y_{322} &=& y_{231} - y_{321} &=& y_{232} - y_{322} \\
                   &=& y_{233} - y_{332} &=& y_{312} - y_{321} &=& y_{313} - y_{331} &=& y_{323} - y_{332} &=& 0,
\end{array}$$
$$\begin{array}{cccccclc}
 y_{332}^2-y_{322}y_{333} &=& y_{331}y_{332}-y_{321}y_{333} &=& y_{322}y_{332}-y_{222}y_{333} &=& y_{321}y_{332}-y_{221}y_{333} &= \\
 y_{331}^2-y_{311}y_{333} &=& y_{311}y_{332}-y_{211}y_{333} &=& y_{322}y_{331}-y_{221}y_{333} &=& y_{321}y_{331}-y_{211}y_{333} &= \\
 y_{322}^2-y_{222}y_{332} &=& y_{311}y_{331}-y_{111}y_{333} &=& y_{222}y_{331}-y_{221}y_{332} &=& y_{221}y_{331}-y_{211}y_{332} &= \\
 y_{321}^2-y_{211}y_{332} &=& y_{211}y_{331}-y_{111}y_{332} &=& y_{321}y_{322}-y_{221}y_{332} &=& y_{311}y_{322}-y_{211}y_{332} &=  \\
 y_{311}^2-y_{111}y_{331} &=& y_{311}y_{321}-y_{111}y_{332} &=& y_{222}y_{321}-y_{221}y_{322} &=& y_{221}y_{321}-y_{211}y_{322} &= \\
 y_{221}^2-y_{211}y_{222} &=& y_{211}y_{321}-y_{111}y_{322} &=& y_{222}y_{311}-y_{211}y_{322} &=& y_{221}y_{311}-y_{111}y_{322} &= \\
 y_{211}^2-y_{111}y_{221} &=& y_{211}y_{311}-y_{111}y_{321} &=& y_{211}y_{221}-y_{111}y_{222} &=& 0.
\end{array}$$
}
A simple verification yields that these equations define an irreducible variety of dimension $3$ and degree $9$,
which must be our eliminant.

\subsubsection{Coupling Constants}\label{s:dp0coupling}
In the above treatment, we have fixed a precise, known, form of the superpotential; essentially, we have fixed the coupling constants therein to be $\epsilon_{\alpha\beta\gamma}$.
Of the 27 possible gauge invariants which could contribute to $W$, we could take arbitrary complex constants to prefix each.
Indeed, we could even have composite loops and thus infinite number of choices.
Here, for convenience, we only consider minimal loops of length~3.

We study the resulting moduli space of the elimination problem \eqref{elim-dp0}
by considering different choices for the $\epsilon_{\alpha\beta\gamma}$.
First, we consider taking each $\epsilon_{\alpha\beta\gamma}$ to be a random complex number.
The numerical irreducible decomposition of the $9$ generators involving
only the $X$ variables yields six irreducible component of dimension $3$,
three sextic and three linear.  Using \eqref{eq:Dim}, each component projects to a zero-dimensional
variety, namely the origin.

Next, we consider taking $\epsilon_{\alpha\alpha\alpha}$ to be any nonzero complex number
and all others to be zero.  Then, the $9$ generators involving only the $X$ variables
define a monomial ideal of dimension $3$ that decomposes into $27$ irreducible linear components.
As above, each component projects to a zero-dimensional variety, namely the origin.

Finally, if we take each $\epsilon_{\alpha\beta\gamma}$ to be $1$, the $9$ generators involving only
the $X$ variables decomposes into $3$ linear components of dimension $7$.
Using \eqref{eq:Dim}, we know that each component projects to a $5$ dimensional variety.
Using the approach of \S~\ref{s:Interp}, one readily obtains that each
of these $5$ dimensional varieties is generated by $15$ linear and $24$ quadratic polynomials,
and can verify that each has degree $12$.  The following are the generators for one of
these varieties with the generators for the
other two obtained by a cyclic permutation on the indices:
{\footnotesize
$$\begin{array}{ccccccc}
y_{111}-y_{212}-y_{213}-y_{312}-y_{313} &=& y_{331}+y_{332}+y_{333} &=& y_{321}+y_{322}+y_{323} &=& \\
y_{121}-y_{222}-y_{223}-y_{322}-y_{323} &=& y_{311}+y_{312}+y_{313} &=& y_{231}+y_{232}+y_{233} &=& \\
y_{131}-y_{232}-y_{233}-y_{332}-y_{333} &=& y_{221}+y_{222}+y_{223} &=& y_{211}+y_{212}+y_{213} &=& \\
y_{133}+y_{233}+y_{333} &=& y_{132}+y_{232}+y_{332} &=& y_{123}+y_{223}+y_{323} &=& \\
y_{122}+y_{222}+y_{322} &=& y_{113}+y_{213}+y_{313} &=& y_{112}+y_{212}+y_{312} &=& 0,
\end{array}$$
$$\begin{array}{ccccccccc}
y_{323}y_{332}-y_{322}y_{333} &=& y_{313}y_{332}-y_{312}y_{333} &=& y_{233}y_{332}-y_{232}y_{333} &=& y_{223}y_{332}-y_{222}y_{333} &=& \\
y_{213}y_{332}-y_{212}y_{333} &=& y_{233}y_{323}-y_{223}y_{333} &=& y_{232}y_{323}-y_{222}y_{333} &=& y_{313}y_{322}-y_{312}y_{323} &=& \\
y_{233}y_{322}-y_{222}y_{333} &=& y_{232}y_{322}-y_{222}y_{332} &=& y_{223}y_{322}-y_{222}y_{323} &=& y_{213}y_{322}-y_{212}y_{323} &=& \\
y_{233}y_{313}-y_{213}y_{333} &=& y_{232}y_{313}-y_{212}y_{333} &=& y_{223}y_{313}-y_{213}y_{323} &=& y_{222}y_{313}-y_{212}y_{323} &=& \\
y_{233}y_{312}-y_{212}y_{333} &=& y_{232}y_{312}-y_{212}y_{332} &=& y_{223}y_{312}-y_{212}y_{323} &=& y_{222}y_{312}-y_{212}y_{322} &=& \\
y_{213}y_{312}-y_{212}y_{313} &=& y_{223}y_{232}-y_{222}y_{233} &=& y_{213}y_{232}-y_{212}y_{233} &=& y_{213}y_{222}-y_{212}y_{223} &=& 0.
\end{array}$$
}
The Hilbert series for each irreducible component of the eliminant is
$$\dfrac{1+7t+4t^2}{(1-t)^5}$$
with the Hilbert series of their union being $(1+15t+18t^2+2t^3)/(1-t)^5$.

\subsection{Instantons on $\IC^2$}
The moduli space of instantons is an important but notoriously difficult object to study, for physicists and mathematicians alike.
Even for flat $\IR^4$ (or, equivalently, the complexified $\IC^2$) as the base space on which the instantons are realized as stable sheaves, the situation is unwieldy.
Beautiful combinatorial approaches have been studied in \cite{Nekrasov:2004vw,Marino:2004cn,Nakajima:2003pg}.
Recently, some nice progress has been made in understanding the situation of $k=1$ and 2 instantons with various gauge groups \cite{Hanany:2012dm,Benvenuti:2010pq}, using Gr\"obner basis techniques as well as direct integration using the Molien formula.
In \cite{Mehta:2012wk}, we alluded to some preliminary results on how one might approach the problem numerically.
In this subsection, let us undertake a more systematic investigation.

For concreteness, let us focus on $k \in \IZ_+$ instantons with gauge group $U(N)$.
In \cite{Benvenuti:2010pq}, it was realized that the instanton moduli space $\cM_{k,N}$ is simply the Higgs branch of the vacuum moduli space of the following $\cN=1$ quiver gauge theory in four dimensions, whereby providing us the perfect formalism as a testing ground:
\begin{equation}
\begin{array}{cc}
\begin{array}{c} \includegraphics[trim= 0mm 0mm 0mm 0mm,clip,width=2.5in]{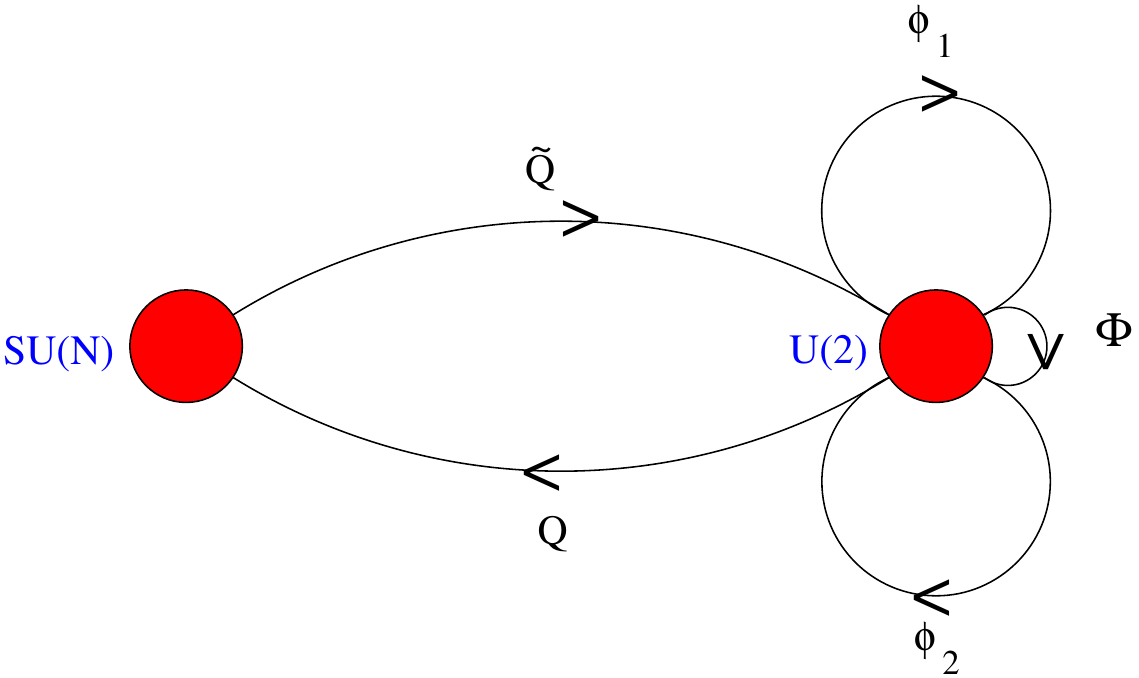} \end{array}
&
W= \sum\limits_{i=1}^N\sum\limits_{a,b=1}^k\tilde{Q}_i^a \Phi^a_b Q^i_b +
   \sum\limits_{\alpha,\beta=1}^2 \epsilon^{\alpha\beta}
   \sum\limits_{a,b,c=1}^k (\phi_\alpha)^a_b \Phi^b_c (\phi_\beta)^c_a
\end{array} \ .
\end{equation}
In the above, we have fields $Q_i^a$ which is a bi-fundamental of $SU(N) \times U(k)$, meaning that it is an $N \times k$ matrix of complex entries, likewise we have $\tilde{Q}^i_a$, which is $k \times N$.
Moreover, we have $(\phi_{\alpha=1,2})^a_b$ which are adjoint fields under the $U(k)$, meaning that they are $k \times k$ complex matrices and $\epsilon^{\alpha \beta}$ is the totally anti-symmetric matrix {\scriptsize $\left(\begin{array}{cc} 0 & 1 \\ -1 & 0 \end{array} \right)$}.
Finally, we also have $\Phi^a_b$ which is another adjoint of $U(k)$.
Throughout, we will use indices $a,b,c,\ldots = 1, \ldots, k$ to associate with $U(k)$, $i,j = 1, \ldots, N$ to associate with $SU(N)$, and $\alpha,\beta=1,2$.

The moduli space in which we are interested is the Higgs branch wherein the vacuum expectation value of the field $\Phi$ vanishes.
Therefore, we have gauge invariant operators composed of the loops other than $\Phi$ in the quiver diagram, which are of the following types:
(1)
$\tr \left[ (\phi_1)^{A} \cdot (\phi_2)^{B} \right]$ where $\cdot$ and $\tr$ are $k \times k$ matrix multiplication and trace, and the matrix exponents $A,B = 0, \ldots, k$ range so that $0 < A + B \le k$;
(2)
$\sum\limits_{i=1}^N\tilde{Q}^i_a Q_i^b$ for $a,b = 1, \ldots, k$;
(3)
$\sum\limits_{a,b=1}^k \tilde{Q}^i_a (\phi_\alpha)^a_b Q^b_j$ for $i,j=1,\ldots,N$ and $\alpha =1,2$.
Next, taking the partial derivatives of $W$ and setting $\Phi$ to zero, the only non-trivial F-term is
$\sum\limits_{i=1}^N Q^i_a \tilde{Q}^b_i + [\phi_1, \phi_2]^b_a$ for $a,b=1,\ldots,k$ with the usual commutator for $k \times k$ matrices.

In summary, we have the following elimination problem.
The indices are such that $i,j = 1, \ldots, N$; $a,b = 1, \ldots, k$;
$\alpha=1,2$;
and $A,B = 0, \ldots, k$ such that $0 < A + B \le k$.
\begin{eqnarray}
\nn
\mbox{ Given ideal } &: &
\left\{
\begin{array}{l}
y^1_{AB} - \tr \left[ (\phi_1)^{A} \cdot (\phi_2)^{B} \right] \ , \\
y^2_{ab} - \sum\limits_{i=1}^N\tilde{Q}^i_a Q_i^b \ , \\
y^3_{ij\alpha} - \sum\limits_{a,b=1}^k \tilde{Q}^i_a (\phi_\alpha)^a_b Q^b_j \ ,\\
\sum\limits_{i=1}^N Q^i_a \tilde{Q}^b_i + [\phi_1, \phi_2]^b_a
\end{array}
\right\} \subset
\IC[y^1_{AB}, y^2_{ab}, y^3_{ij\alpha};
Q_i^a, \tilde{Q}^i_a, (\phi_{\alpha})^a_b
]
\\
\mbox{ Eliminate } &: &
Q_i^a, \ \tilde{Q}^i_a, \ (\phi_{\alpha})^a_b.
\end{eqnarray}
Component-wise, we thus have $2k(N + k)$ variables coming from the quiver fields as well as $3k(k+1)/2 + 2N^2$ coming from the $y$-variables.
Hence we are performing elimination of $2k(N + k)$ variables in
$\IC^{\frac{1}{2} \left(7 k^2+4 k N+3 k+4 N^2\right)}$.

The following describes the cases $(k,N)$ for $k+N \leq 5$.

The case $(k,N) = (1,1)$ is an example that one can solve by hand.  The one generator not involving
the $y$ variables is a monomial that defines two irreducible components, both of which project to the same plane in $\IC^5$
with Hilbert series $1/(1-t)^2$.

For $(k,N) = (1,2)$, it is easy to verify that one generator not involving the $y$ variables
is irreducible.  Thus, the eliminant is also irreducible that one can easily compute
has dimension $4$ and degree $6$ with Hilbert series $(1+4t+t^2)/(1-t)^4$.

For $(k,N) = (1,3)$, the one generator not involving the $y$ variables is irreducible.
In this case, the eliminant is an irreducible variety of dimension $6$ and degree $30$
with Hilbert series $(1+12t+15t^2+2t^3)/(1-t)^6$.

For $(k,N) = (1,4)$, as with the previous two cases, the one generator not involving
the $y$ variables is irreducible.  The eliminant is irreducible of dimension $8$
and degree $140$ with Hilbert series $(1 + 24t + 72t^2 + 40t^3 + 3t^4) / (1-t)^8$.

For $(k,N) = (2,1)$, the given ideal has three irreducible components, all
of dimension $8$.  However, the eliminant only has one irreducible component,
which has dimension $6$ and degree $6$ with Hilbert series $(1+2t+2t^2+t^3)/(1-t)^6$.

For $(k,N) = (2,2)$, the four generators not involving the $y$ variables define
an irreducible variety of dimension $12$ and degree $16$.
With \eqref{eq:Dim}, we have that the eliminant is irreducible of dimension $10$.
Using \sref{s:Slicing}, we find the eliminant has degree $72$ with Hilbert series
$(1+4t+10t^2+19t^3+21t^4+12t^5+4t^6+t^7)/(1-t)^{10}$.

For $(k,N) = (2,3)$, the four generators not involving the $y$ variables define
an irreducible variety of dimension $16$ and degree $16$ with the
eliminant being irreducible of dimension $14$.
Using \sref{s:Slicing}, we find the eliminant has degree $2232$ with Hilbert series
$(1+10t+55t^2+183t^3+410t^4+611t^5+572t^6+306t^7+78t^8+6t^9)/(1-t)^{14}$.

For $(k,N) = (3,1)$, the original ideal has four irreducible components, all
of dimension $15$.  However, the eliminant only has one irreducible component,
which has dimension $10$ and degree $180$ with Hilbert series
$$(1+7t+19t^2+37t^3+52t^4+37t^5+19t^6+7t^7+t^8)/(1-t)^{10}.$$

For $(k,N) = (3,2)$, the nine generators not involving the $y$ variables define
an irreducible variety of dimension $21$ and degree $512$.
The eliminant is irreducible of dimension $18$ and degree $1020$ with Hilbert series
$$(1+5t+15t^2+34t^3+62t^4+96t^5+130t^6+157t^7+170t^8+164t^9+126t^{10}+52t^{11}+8t^{12})/(1-t)^{18}.$$

Finally, for $(k,N) = (4,1)$, the $16$ generators not involving the $y$ variables
define a variety of dimension $24$ and degree $65536$ which decomposes
into $5$ irreducible components.  However, the eliminant is irreducible
of dimension $14$ and degree $11060$ with Hilbert series
$$\begin{array}{l}(1 + 15t + 84t^2 + 300t^3 + 825t^4 + 1809t^5 + 2658t^6 + \\ ~~~~~~~~~~~~~~~~~~~~~~~~~~~~~~~~~~~
2598t^7 + 1767t^8 + 793t^9 + 195t^{10} + 14t^{11} + t^{12})/(1-t)^{14}.\end{array}$$

\subsection{Defining Algebraic Varieties}
It should be noted that our method of elimination applies to situations far beyond merely computing the moduli space of vacua for gauge theories; the setup is indeed applicable to problems in mathematics and physics alike.
In this section, let us take a purely mathematical example which has over the years become important to the study of string compactifications.

The earliest and still one of the most important dataset of Calabi-Yau threefolds, which constitute the corner-stone of compactifications of the ten-dimensional superstring, is the so-called complete intersection Calabi-Yau threefolds, or CICYs for short \cite{Candelas:1987kf,Anderson:2007nc}.
These manifolds are defined as smooth embeddings into products of complex projective spaces $\IP^{n_1} \times \ldots \times \IP^{n_k}$ by polynomials of appropriate multi-degree;
the complete intersection criterion requires that there are $\sum\limits_{i=1}^k n_k - 3$ polynomials.
We will now take some explicit examples from this dataset and investigate some typical questions which arise.
\subsubsection{Embeddings into Projective Space}
Sometimes, we may wish to untangle the structure of the ambient space of product of projective spaces and embed our manifold into a single projective space.
The advantage of doing so is that we have a single grading when dealing with bundles defined on the ambient projective space (which can then be restricted on the Calabi-Yau manifold) whereby facilitating computations of cohomology.
The down-side is that generically we will lose the complete intersection property, though for purposes of computational algebraic geometry this is not particularly troublesome.

Luckily, there is a standard technique of embedding product of projective spaces into a single one: the {\em Segr\`e embedding}.
To illustrate, let us consider the following famous Calabi-Yau manifold, the so-called bi-cubic (we take the form from \cite[\S~2.2]{Candelas:2007ac}),
\begin{equation}\label{bicubic}
\left[
\begin{array}{c|c}
\IP^2_{[x_0:x_1:x_2]} & 3 \\
\IP^2_{[y_0:y_1:y_2]} & 3
\end{array}
\right] \ ,
\qquad
\begin{array}{rcl}
B(x,y) &:=&
x_0x_1x_2 \sum\limits_{i=0}^2 y_i^3 +
y_0y_1y_2 \sum\limits_{i=0}^2 x_i^3 +
x_0x_1x_2y_0y_1y_2 + \\
&& + \sum\limits_{i,j=0}^2 x_i^3 y_{i+j}^3 +
\sum\limits_{i,j=0}^2\sum\limits_{k=\pm1} x_i^2x_{i+k} y_{i+j}^2 y_{i+j+k}
= 0
\end{array}
\end{equation}
The matrix configuration specifies that we have a single polynomial of bi-degree $(3,3)$ in $\IP^2 \times \IP^2$ whose respective projective coordinates we have labeled explicitly.
We have also given an example of this polynomial, as indicated.
The subscripts on the coordinate variables are defined modulo 2.

The Segr\`e embedding here takes pair-wise products of the $x,y$ projective coordinates, thus taking $\IP^2 \times \IP^2$ into $\IP^8$:
\begin{equation}
S : (x_0, x_1, x_2) \times (y_0, y_1, y_2) \longrightarrow
z_{m=0,\ldots,8} := (x_i y_j)_{i,j=0,1,2} \ .
\end{equation}
We wish to compute the defining equation $B$ of the bi-cubic in $\IP^8$.
This, again, is described as an elimination problem:

\begin{eqnarray}
\nn
\mbox{ Given ideal } &: &
\left\{
\begin{array}{c}
z_0 - x_0 y_0 \ , \\
z_1 - x_0 y_1 \ , \\
\vdots\\
z_8 - x_2 y_2 \ , \\
\begin{array}{l}
x_0x_1x_2 \sum\limits_{i=0}^2 y_i^3 +
y_0y_1y_2 \sum\limits_{i=0}^2 x_i^3 +
x_0x_1x_2y_0y_1y_2 + \\
+ \sum\limits_{i,j=0}^2 x_i^3 y_{i+j}^3 +
\sum\limits_{i,j=0}^2\sum\limits_{k=\pm1} x_i^2x_{i+k} y_{i+j}^2 y_{i+j+k}
\end{array}
\end{array}
\right\} \subset \IC[x_{0,1,2}, \ y_{0,1,2}; \ \ z_{0,\ldots,8}]
\\
\mbox{ Eliminate } &: &
x_{0,1,2}, \ y_{0,1,2}
\end{eqnarray}
Hence, we wish to eliminate 6 of the 15 variables.
It is easy to verify that $\Var(B)$ and the eliminant is
irreducible of dimension $3$.  This implies that the eliminant is a cubic hypersurface inside the variety defined
by the nine Pl\"ucker relations:
$$\begin{array}{ccccccccccc}
z_0 z_4 - z_1 z_3 &=& z_0 z_5 - z_2 z_3 &=& z_0 z_7 - z_1 z_6 &=& z_0 z_8 - z_2 z_6 &=& z_1 z_5 - z_2 z_4 &\\
                  &=& z_1 z_8 - z_2 z_7 &=& z_3 z_7 - z_4 z_6 &=& z_3 z_8 - z_5 z_6 &=& z_4 z_8 - z_5 z_7 &=& 0.
\end{array}$$
Since the Pl\"ucker relations define a variety of dimension $4$ and degree $6$ in $\IP^8$,
the eliminant is a dimension $3$ variety of degree $18$ in $\IP^8$.
The approach of \S~\ref{s:LLL} computes the final cubic relation:
\begin{eqnarray*}
&&z_0^3 + z_0 z_1 z_2 + z_0 z_1 z_3 + z_0 z_2 z_6 + z_0 z_3 z_6 + z_1^3 + z_1^2 z_6 + z_1 z_2 z_4 + z_1 z_3 z_4 + z_1 z_4 z_7 \\
&&~~~+ z_1 z_6^2 + z_2^3 + z_2^2 z_3 + z_2^2 z_7 + z_2 z_3^2 + z_2 z_4 z_5 + z_2 z_4 z_6 + z_2 z_5 z_8 + z_2 z_6 z_8 + z_2 z_7^2 \\
&&~~~+ z_3^3 + z_3 z_4 z_5 + z_3 z_4 z_6 + z_4^3 + z_4 z_5 z_7 + z_4 z_6 z_7 + z_5^3 + z_5^2 z_6 + z_5 z_6^2 + z_5 z_7 z_8 + z_6^3 \\
&&~~~+ z_6 z_7 z_8 + z_7^3 + z_8^3 = 0.
\end{eqnarray*}

\subsubsection{Checking Smoothness}
A related problem to the above is whether one could efficiently ensure that our algebraic models for the manifolds are smooth.
In the example of \eqref{bicubic}, we have chosen a specific bi-cubic with fixed complex coefficients.
In general, we can form ${3+3-1 \choose 3-1} = 10$ cubics in 3 variables; thus in general we could have $10 \times 10 = 100$ possible bi-cubic monomials, each of which, together prefixed by some arbitrary coefficient, could potentially contribute to out defining equation $B(x,y)$.
This is, of course, very much analogous to the situation considered in \sref{s:dp0coupling}.
With an overall factor, we thus have a moduli space of projective dimension 99 possible manifolds:
\begin{equation}
\sum\limits_{{\tiny \begin{array}{c}0 \le i,j,k \le 3\\ i+j+k=3\end{array}}}
\sum\limits_{{\tiny \begin{array}{c}0 \le i',j',k' \le 3\\ i'+j'+k'=3\end{array}}}
\
A_{ijk,i'j'k'}
\ \
x_0^ix_1^jx_2^k
y_0^iy_1^jy_2^k
= 0
\end{equation}
We must impose conditions on the coefficients $A_{ijk,i'j'k'}$ such that the above represents a smooth threefold.
Very often, we need to decide this efficiently and over large number of choices of the coefficients.
Traditionally, this has been achieved in the literature by working over a coefficient field being $\IF_p$ for some prime $p$ and checking over several small values of $p$.

We will take a perhaps more gratifying approach and utilize our efficient numerical methods.
Adhering to our example of \eqref{bicubic} for concreteness, suppose we are given
\begin{eqnarray}
\nn
B(x,y; C, D) &:=&
x_0x_1x_2 \sum\limits_{i=0} y_i^3 +
y_0y_1y_2 \sum\limits_{i=0} x_i^3 +
x_0x_1x_2y_0y_1y_2
+ \sum\limits_{i,j=0}^2 C_j x_i^3 y_{i+j}^3 + \\
&& \qquad
+ \sum\limits_{i,j=0}^2\sum\limits_{k=\pm1} D_{j,k} x_i^2x_{i+k} y_{i+j}^2 y_{i+j+k}
= 0
\end{eqnarray}
with 3 complex coefficients $C_{0,1,2}$ and 6 complex coefficients $D_{0,1,2; \pm}$.
Now we wish to check for what choices of these coefficients within our moduli space does the bi-cubic define a smooth Calabi-Yau threefold.
Therefore, we need to check the gradient of partial derivatives with respect to the coordinates and see whether the simultaneous solution of these, together with $B(x,y; C,D)$, indeed is the empty set in $\IP^2 \times \IP^2$.

Once more, therefore, we are confronted with an elimination problem:
\begin{eqnarray}
\nn
\mbox{ Given ideal } &: &
\left\{
\begin{array}{l}
B(x,y; C,D) \ , \\
\frac{\partial}{\partial x_i} B(x,y; C,D) \ , \\
\frac{\partial}{\partial y_i} B(x,y; C,D)
\end{array}
\right\} \subset
\IC[x_{0,1,2}, y_{0,1,2}; \ C_{0,1,2}, D_{0,1,2; \pm}]
\\
\mbox{ Eliminate } &: &
x_{0,1,2}, y_{0,1,2} \ .
\end{eqnarray}

Following \S~\ref{s:GeneralApproach}, we first compute a numerical irreducible decomposition for the given
ideal.  Over $\IC$, this yields $37$ irreducible components of codimension $5$, $36$ of which are linear spaces
and the other has degree $2101$.  Each component projects to a hypersurface in $\IC^9$.
For $\omega = (1 + i \sqrt{3})/2$ and $\overline{\omega} = (1-i\sqrt{3})/2$, each of the following twelve hyperplanes are the image of
exactly $3$ irreducible components:
$$
\begin{array}{l}
C_0 = 0, ~~~ C_1 = 0, ~~~ C_2 = 0, \\
3(C_0 + C_1 + C_2 + D_{0+} + D_{0-} + D_{1+} + D_{1-} + D_{2+} + D_{2-}) + 7 = 0 \\
3(C_0 + C_1 + C_2) - 3\,\omega(D_{0+} + D_{1+} + D_{2+}) - 3\,\overline{\omega}(D_{0-} + D_{1-} + D_{2-}) + 7 = 0 \\
3(C_0 + C_1 + C_2) - 3\,\overline{\omega}(D_{0+} + D_{1+} + D_{2+}) - 3\,\omega(D_{0-} + D_{1-} + D_{2-}) + 7 = 0 \\
3(C_0 + C_1 + C_2) + 3(D_{0+} + D_{0-}) - 3\,\omega(D_{1+} + D_{2-}) - 3\,\overline{\omega}(D_{1-} + D_{2+}) - 2 = 0 \\
3(C_0 + C_1 + C_2) + 3(D_{0+} + D_{0-}) - 3\,\overline{\omega}(D_{1+} + D_{2-}) - 3\,\omega(D_{1-} + D_{2+}) - 2 = 0 \\
3(C_0 + C_1 + C_2) + 3(D_{1+} + D_{1-}) - 3\,\omega(D_{0+} + D_{2-}) - 3\,\overline{\omega}(D_{0-} + D_{2+}) - 2 = 0 \\
3(C_0 + C_1 + C_2) + 3(D_{1+} + D_{1-}) - 3\,\overline{\omega}(D_{0+} + D_{2-}) - 3\,\omega(D_{0-} + D_{2+}) - 2 = 0 \\
3(C_0 + C_1 + C_2) + 3(D_{2+} + D_{2-}) - 3\,\omega(D_{0-} + D_{1-}) - 3\,\overline{\omega}(D_{0+} + D_{1+}) - 2 = 0 \\
3(C_0 + C_1 + C_2) + 3(D_{2+} + D_{2-}) - 3\,\overline{\omega}(D_{0-} + D_{1-}) - 3\,\omega(D_{0+} + D_{1+}) - 2 = 0
\end{array}
$$
The other component projects to a degree $111$ hypersurface in $\IC^9$.
Therefore, outside of these $13$ hypersurfaces,
the bi-cubic defines a smooth Calabi-Yau threefold.

\section{Conclusions and Outlook}
Computational algebraic geometry has been an immensely useful tool to study various theoretical physics problems, more prominently in string and gauge theories. However, the time has come when one needs to transition from the computation of relatively small models (say, number of variables and equations are of order 10)
to the computation of more elaborate models.  Unfortunately, the standard computational algebraic geometry methods run out of the steam
in this transition because the Gr\"obner basis techniques on which all these methods are based are exponential growth in running time and memory consumption.

We have been introducing a novel numerical method called
numerical algebraic geometry (NAG), which efficiently resolves some shortcomings of the
traditional Gr\"obner basis technique including being completely parallelizable, to problems which involve large systems of polynomial equations, arising especially in theoretical physics.

The typical question in gauge and string theory is concerned with the geometrical and particularly topological nature of the solution space, such as the dimension, the number of branches (irreducible components), how many isolated solutions, etc., for this purpose, computational geometry is somewhat an over-kill since we usually assign random values to the parameters in any event and then perform
Gr\"obner reduction.

For example, if we wish to check whether a geometrically engineered gauge theory indeed has a holographic dual being a certain Calabi-Yau manifold, we often set the coupling constants and mass parameters in the Lagrangian to generic values, compute the D- and F-flatness conditions, obtain a large set of polynomial equations, interpret it as an algebraic ideal, and then attempt to find topological quantities such as dimension and Hodge numbers.
Therefore, if an efficient methodology were available to numerically resolve these issues then we can bypass the expensive Gr\"obner basis calculation altogether.
In this case, solutions can be computed to arbitrary accuracy which, in many situations, is sufficient for the application.
However, if exact roots are indeed essential, one can always resort back to Gr\"obner reductions or try exactness recovery methods
such as \cite{RecoveryNumSym}.

The present paper is not only a part of our continuing program but it includes a
new algorithm to compute a numerical irreducible decomposition of an eliminated ideal and (in the aCM case) also compute the important Hilbert series.
We have also succinctly described how one of the most important problems in (supersymmetric) gauge theory, viz., finding the geometry of the vacuum moduli space, can be viewed as problems in elimination theory and applied our aforementioned algorithm to solve many examples ranging from geometrical engineering to instanton moduli spaces.
We have even extended it to answering questions which is common-place is algebraic geometry: when is a given variety non-singular?

The landscape of related open problems which confronts us is vast, and given our current tool we are optimistic in facing them.
For example, the instanton moduli spaces for higher number of instanton number and for more complicated Lie groups are still mysterious. Calculating the decomposition and associated Hilbert series gives a handle on the partition function and counting. We have only demonstrated with some examples and marching toward these in general is an obvious direction.
As another example, we have entered the age of string phenomenology where tremendously large databases of Calabi-Yau manifolds and related geometries are being created; issues such as checking smoothness and computing Hilbert functions are the very bread and butter of the field.
Our numerical techniques should enormously help by bypassing the traditional Gr\"obner reductions.
As a final enticement, the geometry of the vacuum of the MSSM is yet unknown, and we will turn to this important computation in upcoming work.

\section*{Acknowledgements}

JH would like to thank the U.S. National Science Foundation for their support through
DMS-1114336 and DMS-1262428, and North Carolina State University.  DM would like to thank the U.S. Department of Energy
for their support under contract no. DE-FG02-85ER40237. YHH would like to thank the Science and
Technology Facilities Council, UK, for an Advanced Fellowship and grant ST/J00037X/1,
the Chinese Ministry of Education, for a Chang-Jiang Chair Professorship at NanKai University,
the U.S. National Science Foundation for grant CCF-1048082, as well as City University,
London and Merton College, Oxford, for their enduring support.

\bibliographystyle{unsrt}
\addcontentsline{toc}{section}{\refname}

\begin{thebibliography}{}

\bibitem{Gray:2005sr}
J. Gray, Y.-H. He, V. Jejjala, and B.D. Nelson.
\newblock {Vacuum geometry and the search for new physics}.
\newblock {\em Phys.Lett.}, B638:253--257, 2006.

\bibitem{Gray:2006jb}
J. Gray, Y.-H. He, V. Jejjala, and B.D. Nelson.
\newblock {Exploring the vacuum geometry of N=1 gauge theories}.
\newblock {\em Nucl.Phys.}, B750:1--27, 2006.

\bibitem{Hanany:2010vu}
A. Hanany, E.E. Jenkins, A.V. Manohar, and G. Torri.
\newblock {Hilbert Series for Flavor Invariants of the Standard Model}.
\newblock {\em JHEP}, 1103:096, 2011.

\bibitem{comp-book}
Y.-H. He, P. Candelas, A. Hanany, A. Lukas and B. Ovrut, Ed.
\newblock {\em {Computational Algebraic Geometry in String and Gauge Theory}}.
\newblock {Special Issue, Advances in High Energy Physics, Hindawi
  publishing,}, 2012.

\bibitem{Gray:2008yu}
J. Gray, A. Hanany, Y.-H. He, V. Jejjala, and N. Mekareeya.
\newblock {SQCD: A Geometric Apercu}.
\newblock {\em JHEP}, 0805:099, 2008.

\bibitem{Gray:2009fy}
J. Gray.
\newblock {A Simple Introduction to Grobner Basis Methods in String
  Phenomenology}.
\newblock {\em Adv. High Energy Phys.}, 217035, 2011.

\bibitem{CLO:07}
D.A. Cox, J. Little, and D. O'Shea.
\newblock {\em Ideals, Varieties, and Algorithms: An Introduction to
  Computational Algebraic Geometry and Commutative Algebra, 3/e (Undergraduate
  Texts in Mathematics)}.
\newblock Springer-Verlag New York, Inc., Secaucus, NJ, USA, 2007.

\bibitem{Faugere99anew}
J.C. Faug\`{e}re.
\newblock A new efficient algorithm for computing groebner bases (f4).
\newblock In {\em Journal of Pure and Applied Algebra}, pages 75--83. ACM
  Press, 1999.

\bibitem{Faug:02}
J.C. Faug\`{e}re.
\newblock A new efficient algorithm for computing gr\"{o}bner bases without
  reduction to zero (f5).
\newblock In {\em ISSAC '02: Proceedings of the 2002 international symposium on
  Symbolic and algebraic computation}, pages 75--83, New York, NY, USA, 2002.
  ACM.

\bibitem{2005math1111G}
V.P. {Gerdt}.
\newblock {Involutive Algorithms for Computing Groebner Bases}.
\newblock {\em ArXiv Mathematics e-prints}, January 2005.

\bibitem{DGPS}
W.~Decker, G.-M.~Greuel, G.~Pfister, and H.~Sch{\"o}nemann.
\newblock {\sc Singular} {3-1-3} --- {A} computer algebra system for polynomial
  computations.
\newblock 2011.
\newblock http://www.singular.uni-kl.de.

\bibitem{CocoaSystem}
{CoCoA} Team.
\newblock {{\hbox{\rm C\kern-.13em o\kern-.07em C\kern-.13em o\kern-.15em A}}}:
  a system for doing {C}omputations in {C}ommutative {A}lgebra.
\newblock Available at \/ {\tt http://cocoa.dima.unige.it}.

\bibitem{M2}
D.R. Grayson and M.E. Stillman.
\newblock Macaulay2, a software system for research in algebraic geometry.
\newblock Available at {http://www.math.uiuc.edu/Macaulay2/}.

\bibitem{BCP:97}
W. Bosma, J. Cannon, and C. Playoust.
\newblock The magma algebra system i: the user language.
\newblock {\em J. Symb. Comput.}, 24(3-4):235--265, 1997.

\bibitem{Gray:2006gn}
J. Gray, Y.-H. He, and A. Lukas.
\newblock {Algorithmic Algebraic Geometry and Flux Vacua}.
\newblock {\em JHEP}, 0609:031, 2006.

\bibitem{Gray:2008zs}
J. Gray, Y.-H. He, A. Ilderton, and A. Lukas.
\newblock {STRINGVACUA: A Mathematica Package for Studying Vacuum
  Configurations in String Phenomenology}.
\newblock {\em Comput. Phys. Commun.}, 180:107--119, 2009.

\bibitem{Gray:2007yq}
J. Gray, Y.-H. He, A. Ilderton, and A. Lukas.
\newblock {A New Method for Finding Vacua in String Phenomenology}.
\newblock {\em JHEP}, 0707:023, 2007.

\bibitem{BHSW06}
D.J. Bates, J.D. Hauenstein, A.J. Sommese, and C.W. Wampler.
\newblock Bertini: Software for numerical algebraic geometry.
\newblock Available at www.nd.edu/$\sim$sommese/bertini.

\bibitem{Ver:99}
J. Verschelde.
\newblock Algorithm 795: Phcpack: a general-purpose solver for polynomial
  systems by homotopy continuation.
\newblock {\em ACM Trans. Math. Soft.}, 25(2):251--276, 1999.

\bibitem{GKKTFM:04}
T. Gunji, S. Kim, M. Kojima, A. Takeda, K. Fujisawa, and T. Mizutani.
\newblock Phom: a polyhedral homotopy continuation method for polynomial
  systems.
\newblock {\em Computing}, 73(1):57--77, 2004.

\bibitem{MSW:89}
A.P. Morgan, A.J. Sommese, and L.T. Watson.
\newblock Finding all isolated solutions to polynomial systems using hompack.
\newblock {\em ACM Trans. Math. Softw.}, 15(2):93--122, 1989.

\bibitem{GLW:05}
T. Gao, T.Y. Li, and M. Wu.
\newblock Algorithm 846: Mixedvol: a software package for mixed-volume
  computation.
\newblock {\em ACM Trans. Math. Softw.}, 31(4):555--560, 2005.

\bibitem{Li:03}
T.L. Lee, T.Y. Li, and C.H. Tsai.
\newblock Hom4ps-2.0, a software package for solving polynomial systems by the
  polyhedral homotopy continuation method.
\newblock {\em Computing}, 83:109--133, 2008.

\bibitem{Mehta:2009}
D. Mehta.
\newblock {Lattice vs. Continuum: Landau Gauge Fixing and 't Hooft-Polyakov
  Monopoles}.
\newblock {\em Ph.D. Thesis, The Uni. of Adelaide, Australasian Digital Theses
  Program}, 2009.

\bibitem{Mehta:2009zv}
D. Mehta, A. Sternbeck, L. von Smekal, and A.G. Williams.
\newblock {Lattice Landau Gauge and Algebraic Geometry}.
\newblock {\em PoS}, QCD-TNT09:025, 2009.

\bibitem{Hughes:2012hg}
C. Hughes, D. Mehta, and J.-I. Skullerud. Enumerating Gribov copies on the lattice. arXiv:1203.4847 [hep-lat].

\bibitem{Mehta:2011xs}
D. Mehta.
\newblock {Finding All the Stationary Points of a Potential Energy Landscape
  via Numerical Polynomial Homotopy Continuation Method}.
\newblock {\em Phys. Rev.}, E84:025702, 2011.

\bibitem{Kastner:2011zz}
M. Kastner and D. Mehta.
\newblock {Phase Transitions Detached from Stationary Points of the Energy
  Landscape}.
\newblock {\em Phys. Rev. Lett.}, 107:160602, 2011.

\bibitem{PhysRevE.85.061103}
D. Mehta, J.D. Hauenstein, and M. Kastner.
\newblock Energy-landscape analysis of the two-dimensional nearest-neighbor
  $\phi^{4}$ model.
\newblock {\em Phys. Rev. E}, 85:061103, 2012.

\bibitem{Nerattini:2012pi}
R. Nerattini, M. Kastner, D. Mehta, and L. Casetti.
\newblock {Exploring the energy landscape of XY models.}.
\newblock {arXiv:1211.4800 [cond-mat.stat-mech].}

\bibitem{Maniatis:2012ex}
M. Maniatis and D. Mehta.
\newblock {Minimizing Higgs Potentials via Numerical Polynomial Homotopy
  Continuation}.
\newblock {\em Eur. Phys. J. Plus}, 127:91, 2012.

\bibitem{CamargoMolina:2012hv} 
J.~E.~Camargo-Molina, B.~O'Leary, W.~Porod and F.~Staub.
The Stability Of R-Parity In Supersymmetric Models Extended By $U(1)_{B-L}$.
arXiv:1212.4146 [hep-ph].

\bibitem{Mehta:2011wj}
D. Mehta.
\newblock {Numerical Polynomial Homotopy Continuation Method and String Vacua}.
\newblock {\em Adv. High Energy Phys.}, 2011:263937, 2011.

\bibitem{Mehta:2012wk}
D. Mehta, Y.-H. He, and J.D. Hauenstein.
\newblock {Numerical Algebraic Geometry: A New Perspective on String and Gauge
  Theories}.
\newblock {\em JHEP}, 1207:018, 2012.

\bibitem{He:2013yk} 
Y.~-H.~He, D.~Mehta, M.~Niemerg, M.~Rummel and A.~Valeanu. Exploring the Potential Energy Landscape Over a Large Parameter-Space.
arXiv:1301.0946 [hep-th].
  
\bibitem{MartinezPedrera:2012rs} 
D.~Martinez-Pedrera, D.~Mehta, M.~Rummel and A.~Westphal. Finding all flux vacua in an explicit example.
arXiv:1212.4530 [hep-th].

\bibitem{Projection}
J.D. Hauenstein and A.J. Sommese.
\newblock Witness sets of projections.
\newblock {\em Appl. Math. Comput.}, 217(7):3349--3354, 2010.

\bibitem{ProjectMembership}
J.D. Hauenstein and A.J. Sommese.
\newblock Membership tests for images of algebraic sets by linear projections. Preprint, 2012. Available at www4.ncsu.edu/$\sim$jdhauens/preprintsf.

\bibitem{Migliore}
J.C. Migliore.
\newblock {\em Introduction to liaison theory and deficiency modules}, volume
  165 of {\em Progress in Mathematics}.
\newblock Birkh\"auser Boston Inc., Boston, MA, 1998.

\bibitem{wess1992supersymmetry}
J.~Wess and J.~Bagger.
\newblock {\em Supersymmetry and Supergravity}.
\newblock Princeton Series in Physics. Princeton University Press, 1992.

\bibitem{Luty:1995sd}
M.A. Luty and W. Taylor.
\newblock {Varieties of vacua in classical supersymmetric gauge theories}.
\newblock {\em Phys. Rev.}, D53:3399--3405, 1996.

\bibitem{argyres}
P.C. Argyres.
An Introduction to Global Supersymmetry. Lecture notes (2001).

\bibitem{Buccella:1982nx}
F.~Buccella, J.P. Derendinger, S.~Ferrara, and C.A. Savoy.
\newblock {Patterns of Symmetry Breaking in Supersymmetric Gauge Theories}.
\newblock {\em Phys. Lett.}, B115:375, 1982.

\bibitem{Gatto:1986bt}
R. Gatto and G.~Sartori.
\newblock {Consequences of the complex character of the internal symmetry in
  supersymmetric theories}.
\newblock {\em Commun. Math. Phys.}, 109:327, 1987.

\bibitem{Procesi:1985hr}
C.~Procesi and G.W. Schwarz.
\newblock {The Geometry of Orbit Spaces and Guage Symmetry Breaking in
  Supersymmetric Guage Theories}.
\newblock {\em Phys. Lett.}, B161:117--121, 1985.

\bibitem{Witten:1993yc}
E. Witten.
\newblock {Phases of N=2 theories in two-dimensions}.
\newblock {\em Nucl. Phys.}, B403:159--222, 1993.

\bibitem{GBBIB726}
G.~Greuel and G.~Pfister.
\newblock {\em A Singular Introduction to Commutative Algebra}.
\newblock Springer, 2002.

\bibitem{EliminationMethods}
D.~Wang.
\newblock {\em Elimination methods}.
\newblock Texts and Monographs in Symbolic Computation. Springer-Verlag,
  Vienna, 2001.

\bibitem{SW:05}
A.J. Sommese and C.W. Wampler.
\newblock {\em The numerical solution of systems of polynomials arising in
  Engineering and Science}.
\newblock World Scientific Publishing Company, 2005.

\bibitem{Isosingular}
J.D. Hauenstein and C.W. Wampler.
\newblock Isosingular sets and deflation.
\newblock Preprint, 2011. Available at www4.ncsu.edu/$\sim$jdhauens/preprints.

\bibitem{HilbertZero}
Z.A. Griffin, J.D. Hauenstein, C.~Peterson, and A.J. Sommese.
\newblock Numerical computation of the hilbert function of a zero-scheme.
\newblock Preprint, 2011. Available at www4.ncsu.edu/$\sim$jdhauens/preprints.

\bibitem{RecoveryNumSym}
D.J. Bates, J.D. Hauenstein, T.M. McCoy, C.~Peterson, and A.J. Sommese.
\newblock Recovering exact results from inexact numerical data in algebraic
  geometry.
\newblock Preprint, 2011. Available at www4.ncsu.edu/$\sim$jdhauens/preprints.

\bibitem{LLL}
A.~K. Lenstra, H.~W. Lenstra, Jr., and L.~Lov{\'a}sz.
\newblock Factoring polynomials with rational coefficients.
\newblock {\em Math. Ann.}, 261(4):515--534, 1982.

\bibitem{PSLQ}
H.~Ferguson and D.~Bailey.
\newblock A polynomial time, numerically stable integer relation algorithm.
\newblock Technical report, 1991.

\bibitem{HSW10}
J.D. Hauenstein, A.J. Sommese, and C.W. Wampler.
\newblock Regeneration homotopies for solving systems of polynomials.
\newblock {\em Math. Comp.}, 80(273):345--377, 2011.

\bibitem{Nekrasov:2004vw}
N. Nekrasov and S. Shadchin.
\newblock {ABCD of instantons}.
\newblock {\em Commun. Math. Phys.}, 252:359--391, 2004.

\bibitem{Marino:2004cn}
M. Marino and N. Wyllard.
\newblock {A Note on instanton counting for N=2 gauge theories with classical
  gauge groups}.
\newblock {\em JHEP}, 0405:021, 2004.

\bibitem{Nakajima:2003pg}
H.~Nakajima and K.~Yoshida.
\newblock {Instanton counting on blowup. I}.
\newblock {\em math/0306198}.

\bibitem{Hanany:2012dm}
A. Hanany, N. Mekareeya, and S.S. Razamat.
\newblock {Hilbert Series for Moduli Spaces of Two Instantons}.
\newblock 2012.

\bibitem{Benvenuti:2010pq}
S. Benvenuti, A. Hanany, and N. Mekareeya.
\newblock {The Hilbert Series of the One Instanton Moduli Space}.
\newblock {\em JHEP}, 1006:100, 2010.

\bibitem{Candelas:1987kf}
P.~Candelas, A.M. Dale, C.A. Lutken, and R.~Schimmrigk.
\newblock {Complete Intersection Calabi-Yau Manifolds}.
\newblock {\em Nucl. Phys.}, B298:493, 1988.

\bibitem{Anderson:2007nc}
L.B. Anderson, Y.-H. He, and A. Lukas.
\newblock {Heterotic Compactification, An Algorithmic Approach}.
\newblock {\em JHEP}, 0707:049, 2007.

\bibitem{Candelas:2007ac}
P. Candelas, X. de~la Ossa, Y.-H. He, and B. Szendroi.
\newblock {Triadophilia: A Special Corner in the Landscape}.
\newblock {\em Adv. Theor. Math. Phys.}, 12:2, 2008.


\end{thebibliography}

\end{document}